\documentclass[%
 reprint,
 twocolumn,
 amsmath,amssymb,
aip,jcp,
]{revtex4-2}

\usepackage{natbib}
\usepackage{graphicx}
\usepackage{dcolumn}
\usepackage{bm}
\usepackage{xcolor}
\usepackage{physics}
\usepackage{lipsum}
\usepackage{placeins}

\begin{document}
\title{The Role of Superlattice Phonons in Charge Localization Across Quantum Dot Arrays}
\author{Bokang Hou}
\affiliation {Department of Chemistry, University of
California, Berkeley, California 94720, United States;}
\author{Matthew Coley-O'Rourke}
\affiliation {Department of Chemistry, University of
California, Berkeley, California 94720, United States;}
\author{Uri Banin}
\affiliation {Institute of Chemistry and the Center for
Nanoscience and Nanotechnology, The Hebrew University of
Jerusalem, 91904 Jerusalem, Israel;}
\author{Michael Thoss}
\affiliation {Institute of Physics, Albert-Ludwigs University Freiburg, Hermann-Herder-Straße 3, 79104 Freiburg, Germany;}
\author{Eran Rabani}
\affiliation {Department of Chemistry, University of
California, Berkeley, California 94720, United States;}
\affiliation {Materials Sciences Division, Lawrence Berkeley National Laboratory, Berkeley, California 94720, United States;}
\affiliation {The Raymond and Beverly Sackler Center of Computational Molecular and Materials Science, Tel Aviv University, Tel Aviv 69978, Israel;}

\begin{abstract}
\textbf{Abstract:} Understanding charge transport in semiconductor quantum dot (QD) assemblies is important for developing the next generation of solar cells and light-harvesting devices based on QD technology. One of the key factors that governs the transport in such systems is related to the hybridization between the QDs. Recent experiments have successfully synthesized QD molecules, arrays, and assemblies by directly fusing the QDs, with enhanced hybridization leading to high carrier mobilities and coherent band-like electronic transport. In this work, we theoretically investigate the electron transfer dynamics across a finite CdSe-CdS core-shell QD array, considering up to seven interconnected QDs in one dimension. We find that, even in the absence of structural and size disorder, electron transfer can become localized by the emergent low-frequency superlattice vibrational modes when the connecting neck between QDs is narrow. On the other hand, we also identify a regime where the same vibrational modes facilitate coherent electron transport when the connecting necks are wide. Overall, we elucidate the crucial effects of electronic and superlattice symmetries and their couplings when designing high-mobility devices based on QD superlattices.
\end{abstract}

\maketitle

Nanomaterials offer unique optoelectronic properties for device design in photovoltaics, imaging, and light-emitting diodes.~\cite{kirmani_colloidal_2020,garcia_de_arquer_semiconductor_2021,dai_solution-processed_2014} Many of those devices require an assembly of highly programmable nanoscale building blocks, whose size, shape, and composition can be easily modified.\cite{bassani_nanocrystal_2024} Colloidal semiconductor quantum dots (QDs) represent such tunable building blocks. Experiments have realized assembling QD monomers in different dimensions, from 1D QD arrays \cite{borsoi_shared_2024,braakman_long-distance_2013} to 2D QD monolayers,~\cite{talapin_pbse_2005,kovalenko_colloidal_2009,ondry_colloidal_2021} and 3D superlattice structures.~\cite{murray_self-organization_1995, choi_controlling_2011, abelson_collective_2020} Building those novel nanostructures opens up new ways to design mesoscale materials and devices in a controlled and programmable way. 

\begin{figure}[t!]
    \centering
    {{\includegraphics[width=8cm]{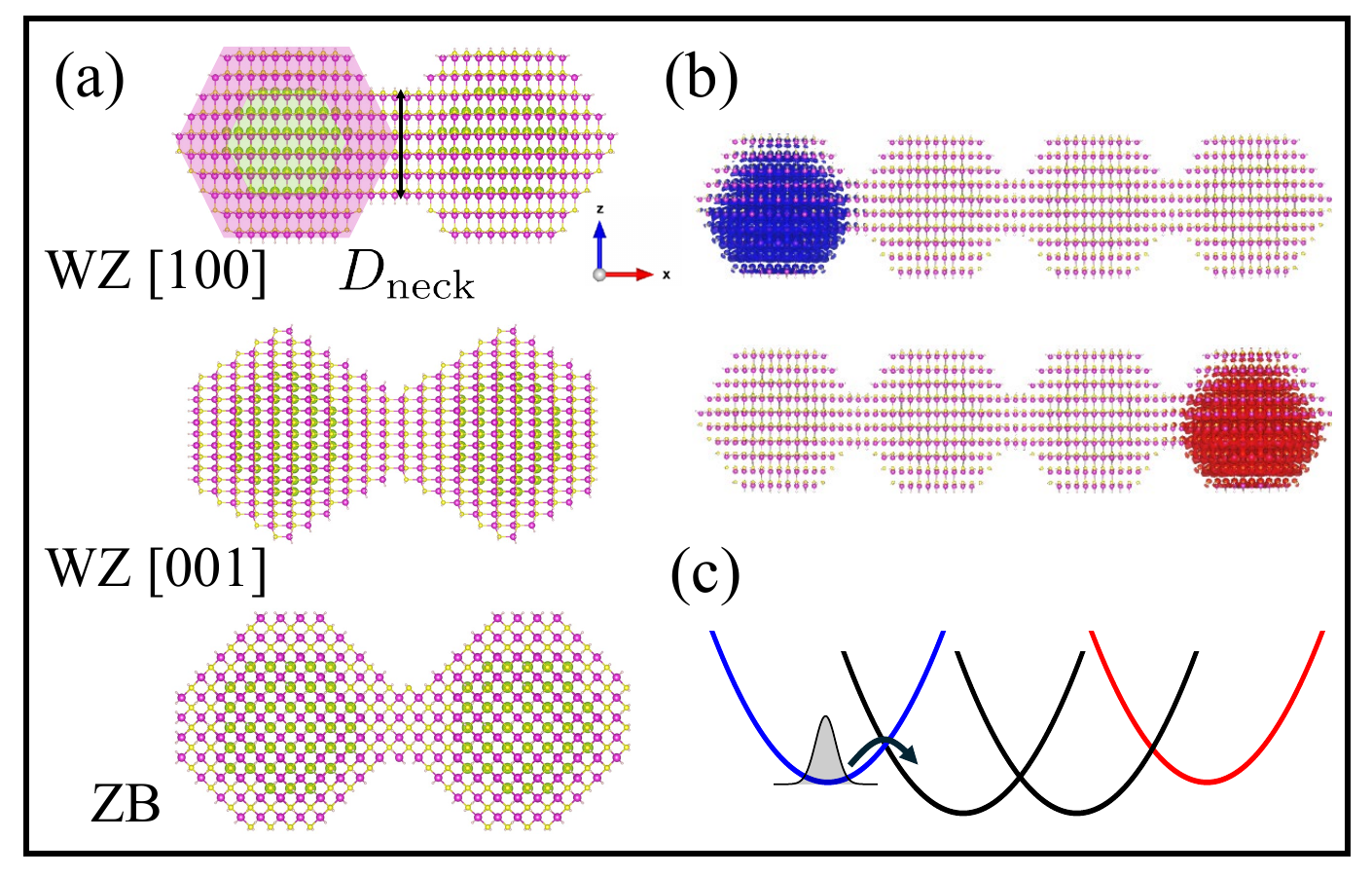} }}
    \caption{(a) QD dimers with [100] and [001] attachment orientations in the wurtzite (WZ) and zincblende (ZB) crystal structures. Green and magenta shaded regions represent the CdSe core and CdS shell respectively. $D_{\rm neck}$ is the neck width. (b) Localized electronic densities of the donor (blue) and acceptor (red) states for a WZ QD tetramer in the [100] attachment. (c) Schematic diagram of the diabatic potential energy surfaces for electron transfer process through a tetramer QD chain.}
    \label{Fig_intro}
\end{figure}

A key design principle of QD assemblies is to achieve high electron mobility across the superlattice, thereby enhancing the performance of QD-based devices such as photovoltaics and photodetectors.\cite{yu_n-type_2003,kagan_charge_2015} Traditionally, the QD assemblies are self-assembled in the solution phase with each QD encapsulated by organic ligands. The organic ligands then form a large energy barrier, making the electrons hop from one site to another incoherently.\cite{yazdani_nanocrystal_2019,yazdani_charge_2020} This process is well described by Marcus theory~\cite{marcus_theory_1956} in which the electron transfer rate is proportional to the square of electronic couplings. Due to the small wavefunction overlap caused by ligands, inhomogeneity, or heterostructures, the transfer time between two QDs is typically in the nanosecond range.\cite{scholes_energetics_2007, xing_effect_2022} Changing organic ligands to inorganic metal chalcogenide ligands has been shown to enhance electron mobility.\cite{kovalenko_colloidal_2009}

Recently, coupled QD structures have been fabricated in the colloidal phase or by epitaxial growth.\cite{cui_colloidal_2019,whitham_charge_2016,pinna_approaching_2023,septianto_enabling_2023} QDs in those structures are interconnected by crystalline bridges without separation by organic ligands. Experiments show that the superlattice structure can be highly ordered with a controllable lattice structure, QD shapes, and connection areas.\cite{abelson_collective_2020,cui_neck_2021,cui_semiconductor_2021} In particular, the connecting neck between QDs is an important parameter that could be controlled experimentally. The neck girth in QD dimers, (termed coupled QD molecules, CQDMs),  enhances the hybridization energies and the red shift of the band gap.\cite{verbitsky_hybridization_2022,levi_effect_2023} It also affects the nature of biexcitons from localized in weakly coupled dimers, to delocalized in the strong coupling regime.\cite{koley_photon_2022,frenkel_two_2023} Strong electronic coupling enabled by a full connecting neck is also important for facile color switching under the application of an electric field in these systems.\cite{ossia_electric-field-induced_2023} In this strong coupling limit, mini-bands form as in bulk material, and wavefunctions delocalize among several QDs as the inter-dot electronic coupling increases significantly.\cite{lazarenkova_electron_2002,kalesaki_electronic_2013,whitham_charge_2016,kavrik_emergence_2022} The transport properties are also enhanced, as shown by the band-like transport in the carrier mobility measurement.\cite{lee_band-like_2011,choi_bandlike_2012,shabaev_dark_2013,lan2020quantum,septianto_enabling_2023}

There are several factors that could affect electron transfer in coupled QD systems. Whitham \textit{et al.}\cite{whitham_charge_2016} showed that the energy disorder due to the inhomogeneity of the QD sizes could cause Anderson-type localization of the electronic wavefunction. Similarly, the variation of the epitaxial connection width (or neck) could also affect the wavefunction overlap between QDs. Other factors, such as defect states, could form electronic traps and decrease electron mobility.\cite{ondry_resilient_2019} Despite numerous studies exploring the effects of structural disorder on charge localization, there is little discussion on the intrinsic factor governing electron transfer properties: electron-phonon coupling.  In a recent study, we developed an atomistic theory to describe electron transfer in coupled QD dimers.\cite{hou_incoherent_2023} Simulations of our model revealed the influence of inter-dot stretching and torsional modes on electron transfer. We also uncovered an intriguing transition from the Marcus-like nonadiabatic electron transfer limit to a coherent adiabatic limit, where rapid electron transfer is governed by dephasing. This transition is controlled by modifying the neck width connecting the two QDs or the shell thicknesses of each QD. One of the most significant questions arising from our previous study is whether the coherent dynamics observed in the adiabatic limit persists beyond the limit of just two QDs and what will be the coherent length scale for transport in a QD assembly.

In this letter, we address these questions in an array of CdSe-CdS core-shell colloidal QDs epitaxially connected in one dimension. We use the semiempirical pseudopotential model to represent both the excess electron states and the coupling of the electron to the nuclear vibration, including the interaction of both superlattice and intra-dot phonons.\cite{jasrasaria2022simulations,hou_incoherent_2023} We explicitly considered the effects of lattice symmetry, attachment orientation, and neck width $D_{\rm neck}$ (see Figure~\ref{Fig_intro}(a) and Supplementary Figure~1) on the electron transfer timescale as a function of the length of the QD array. We observe that the symmetry of the electronic wavefunction influences the interaction between the electron and the superlattice bending and stretching modes. This interaction can either localize or facilitate electron transfer depending on the transfer regime. To explain our findings, we crafted a simplified chain model and derived scaling laws for the primary mode frequency, the electron-phonon coupling, and the reorganization energy, in relation to the number of QD assembly units.

\section{Results and Discussion}

\begin{figure*}[ht!]
    \centering
    {{\includegraphics[width=17.5cm]{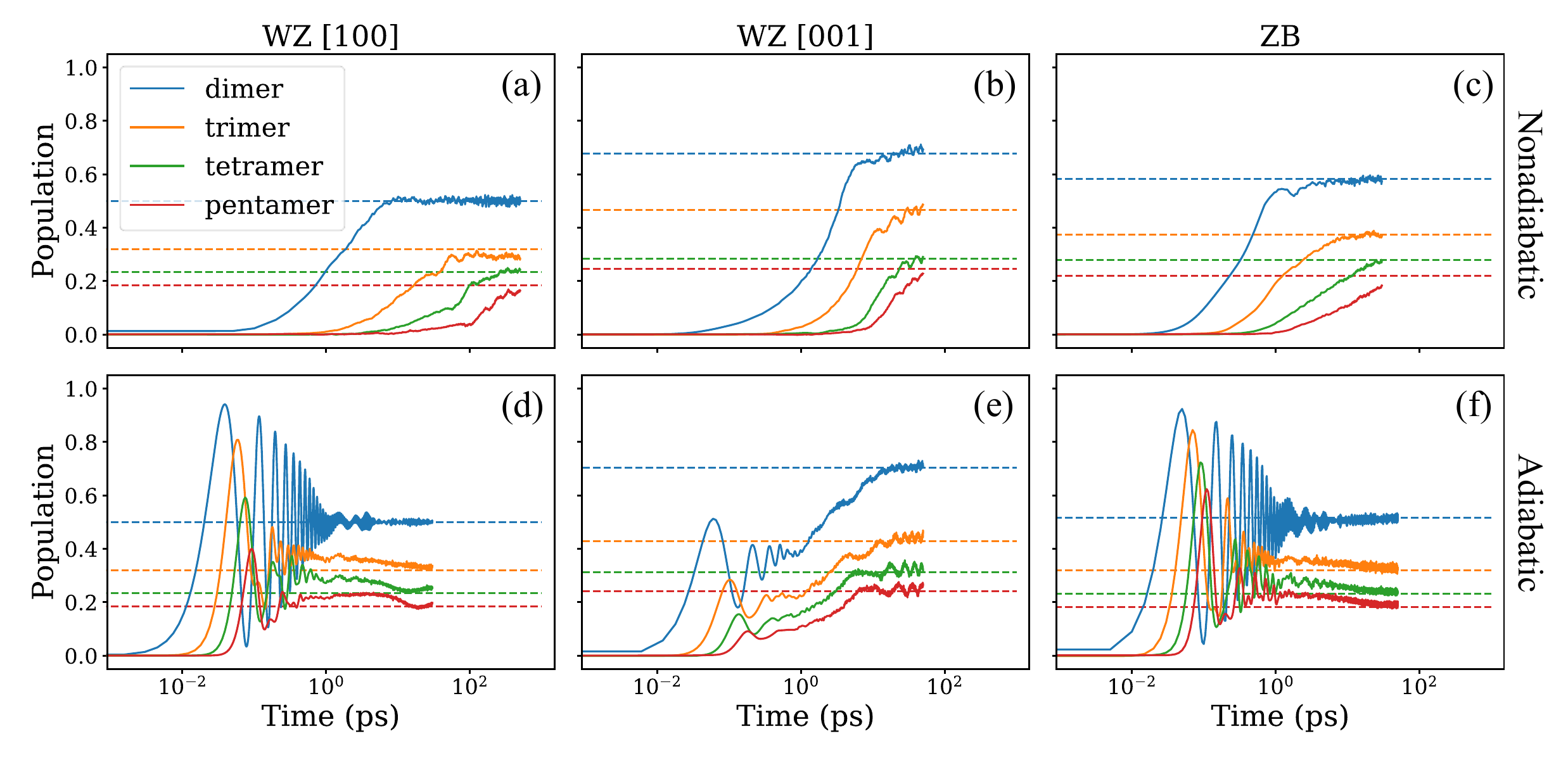} }}
    \caption{Acceptor population dynamics in the narrow-neck nonadiabatic (panels (a)-(c)) and wide-neck adiabatic (panels (d)-(f))) limits, for the WZ [100] attachment (left panels), WZ [001] attachment (middle panels), and ZB attachment (right panels). QD arrays with 2 to 5 units were shown. The equilibrium populations $\mathcal{P}_{\rm eq}$ are indicated by the dashed lines. The dynamics were generated at $T=300$~K.}
    \label{Fig_pop}
\end{figure*}
\textbf{Electron Transfer Model in QD Arrays.} To describe the electron transfer dynamics, we begin by introducing a model Hamiltonian given by a sum of three terms:\cite{hou_incoherent_2023} $H_{\rm QD} = H_{\rm el} + H_{\rm nu} + H_{\rm el-nu}$, where $H_{\rm el}$ describes the electronic system of the QD array, $H_{\rm nu}$ represents the nuclear Hamiltonian, and $H_{\rm el-nu}$ accounts for the interaction between the electronic and nuclear degrees of freedom, approximated to the first order in the nuclear coordinates. More explicitly,
\begin{equation}\label{eq:H_model}
\begin{aligned}
    &H_{\rm el} = \sum_{n}^{N_{\rm el}} \varepsilon_{n}\ket{\phi_n}\bra{\phi_n} + \sum_{m\ne n}^{N_{\rm el}} J_{nm} \ket{\phi_n}\bra{\phi_m} \\
    &H_{\rm nu} = \sum_{\alpha}^{N_{\rm vib}}\frac{1}{2}{P}_{\alpha}^2 + \frac{1}{2}\omega_\alpha^2Q_\alpha^2 \\
    &H_{\rm el-nu} =\sum_{\alpha}^{N_{\rm vib}}\sum_{n,m}^{N_{\rm el}}\ket{\phi_n}\bra{\phi_m}V_{nm}^\alpha{Q}_{\alpha}.
\end{aligned}
\end{equation} 
In the above, $\varepsilon_n$ is the on-site energy of electronic state $\ket{\phi_n}$ at $n$-th QD and $J_{nm}$ is the hybridization energy between the states $\ket{\phi_n}$ and $\ket{\phi_m}$. We used localized quasi-electron states~\cite{foster_canonical_1960} that were obtained from the semi-empirical pseudopotential Hamiltonian $h_{\rm e}$,\cite{hou_incoherent_2023} as schematically illustrated in Figure~\ref{Fig_intro}(b) for the donor and acceptor QDs. The onsite energy and hybridization can then be expressed as $\varepsilon_n= \bra{\phi_n}{h}_{\rm e}\ket{\phi_n}$ and $J_{nm}=\bra{\phi_n}{h}_{\rm e}\ket{\phi_m}$. The nuclear degrees of freedom were described by the Stillinger-Weber force field, with $P_\alpha$ and $Q_\alpha$ denoting the mass-weighted vibrational normal mode coordinates labelled by $\alpha$. These were calculated by diagonalizing the Hessian matrix at the equilibrium geometry of the QD array. The electron-phonon coupling strength between electron states $\ket{\phi_n}$ and $\ket{\phi_m}$ for mode $\alpha$, denoted by $V_{nm}^\alpha$, is obtained from the first-order derivative of the pseudopotential Hamiltonian with respect to the nuclear coordinates.\cite{jasrasaria2022simulations,jasrasaria_interplay_2021} See Methods section and Supplementary Information for details.

Since a full quantum mechanical treatment of the dynamics is still elusive for the current model due to the large number of degrees of freedom, we employed a mixed quantum-classical approach based on spin mapping.\cite{runeson_spin-mapping_2019,runeson_generalized_2020} This approximation is appropriate when the nuclei mass is large and the motion is sluggish, which is the case for CdSe-CdS core-shell QDs. The equations of motion for the mapping variables are given by:
\begin{equation}
\begin{aligned}
\dot{q}_n & = \frac{1}{\hbar}\sum_m^{N_{\rm el}} (\varepsilon_n\delta_{nm} + J_{nm} + \sum_\alpha V_{nm}^\alpha Q_\alpha) p_m, \\
\dot{p}_n & = -\frac{1}{\hbar}\sum_m^{N_{\rm el}} (\varepsilon_n\delta_{nm} + J_{nm} + \sum_\alpha V_{nm}^\alpha Q_\alpha) q_m, \\
\dot{Q}_\alpha & = P_\alpha, \\
\dot{P}_\alpha & = -\omega_\alpha^2Q_\alpha \\
& \quad - \frac{1}{2\hbar}\sum_{n, m}^{N_{\rm el}} V_{nm}^\alpha \left( q_n q_m + p_n p_m - \hbar \gamma_s \delta_{nm} \right).
\end{aligned}
\end{equation}
In the above, $q_n$ and $p_n$ are a pair of mapping variables~\cite{meyera_classical_1979,thoss_mapping_1999} where the population at QD site $n$ can be expressed as an average of $\mathcal{P}_n(t) = \frac{1}{2\hbar}(p_n^2+q_n^2-\hbar\gamma_s)$, with $\gamma_s =  2(\sqrt{N_{\rm el}+1}-1)/N_{\rm el}$ being the zero-point energy correction.\cite{runeson_spin-mapping_2019, runeson_generalized_2020} This choice of $\gamma_s$ also ensures that the system populations approximately relax to the correct thermal equilibrium (for a detailed description of the dynamical method see Supplementary Information section III).\cite{amati_detailed_2023} We assume that the electron is prepared in the lowest donor state with a thermal distribution of the nuclear degrees of freedom. In Figure~\ref{Fig_intro}(c), we plot a schematic diabatic potential energy surface along the QD array, labeling the donor and acceptor states in blue and red, respectively.

\begin{figure*}[ht!]
    \centering
    {{\includegraphics[width=17.5cm]{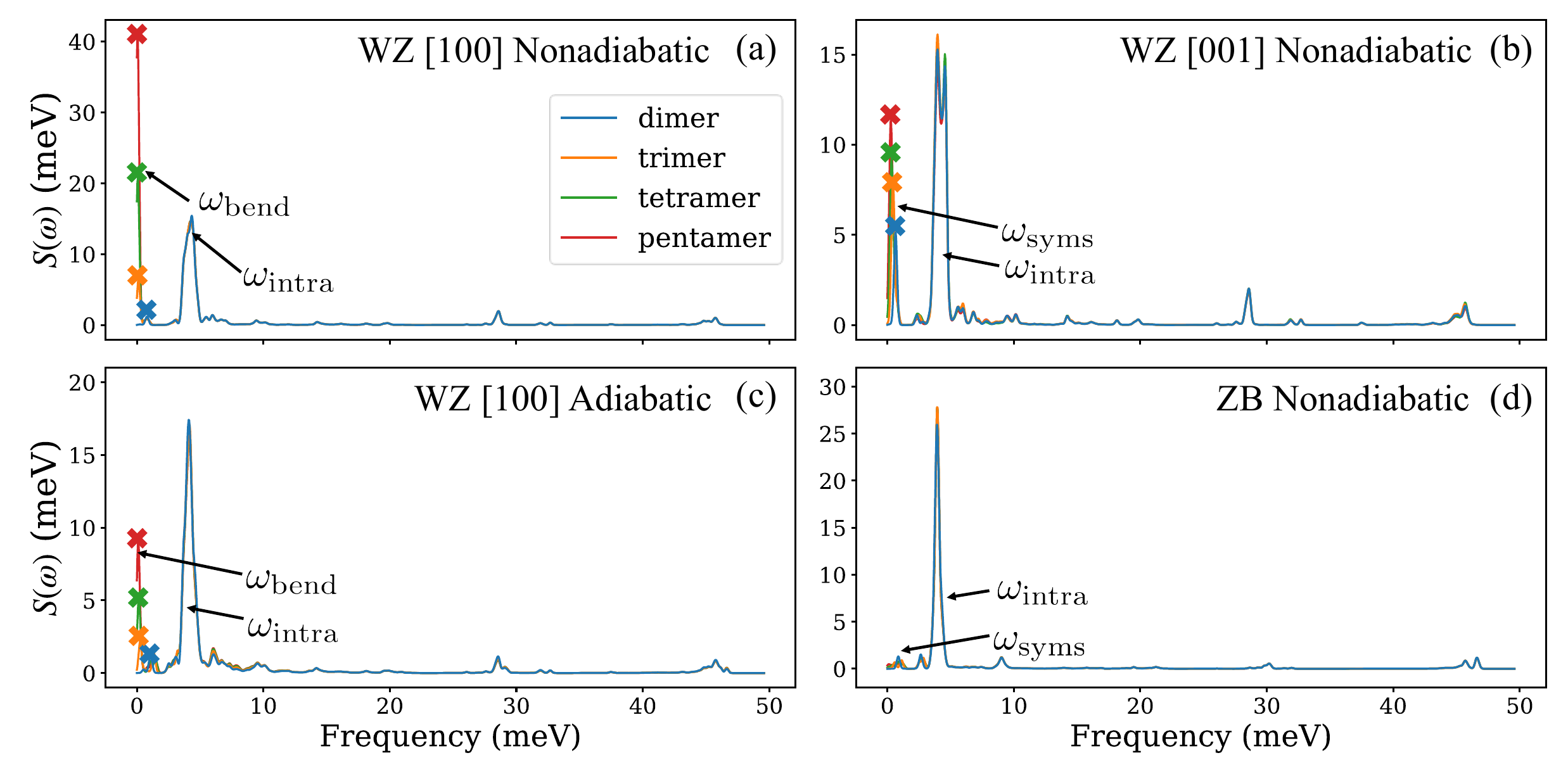} }}
    \caption{Plots of the donor ground state spectral density $S_{11}(\omega)$ for (a) WZ [100] and (b) WZ [001] attachments in the nonadiabatic limit, (c) WZ [100] attachment in the adiabatic limit, and (d) ZB structure in the nonadiabatic limit. Superlattice phonons that significantly contribute to the spectral density are labeled as $\omega_{\rm bend}$ (bending mode), $\omega_{\rm syms}$ (symmetric stretch), and $\omega_{\rm intra}$ (torsional modes).}
    \label{Fig_Jw}
\end{figure*}

QDs with wurtzite (WZ) or zincblende (ZB) lattice structures were considered, and the electron transfer dynamics were studied as a function of the neck dimensions connecting the QDs and the length of the QD chain. For small neck widths, electron transfer is driven by weak electronic hybridization and stronger coupling to fast-moving nuclei (relative to electrons), corresponding to the nonadiabatic regime.\cite{hou_incoherent_2023}  Conversely, for large neck widths and strong hybridization, the dynamics correspond to the adiabatic limit with strong electronic coupling, well-separated potential energy surfaces, and slow lattice fluctuations.\cite{hou_incoherent_2023} Figure~\ref{Fig_pop} plots the dynamics of the acceptor population for WZ and ZB QDs in the nonadiabatic and adiabatic limits.  The transfer reaches equilibrium (shown by the dashed lines) on a time scale of $\approx 1$ to $\approx 100$ picoseconds. The population dynamics shown in Figure~\ref{Fig_pop}, panels (a)-(c) are characterized by overdamped dynamics and the transfer time increases with the QD chain units. The transfer time slows down by nearly two orders of magnitude (depending on the attachment orientation) as the length of the QD chain increases from $2$ to $5$ in the [100] nonadiabatic regime. 

\begin{figure}[ht!]
    \centering
    {{\includegraphics[width=8cm]{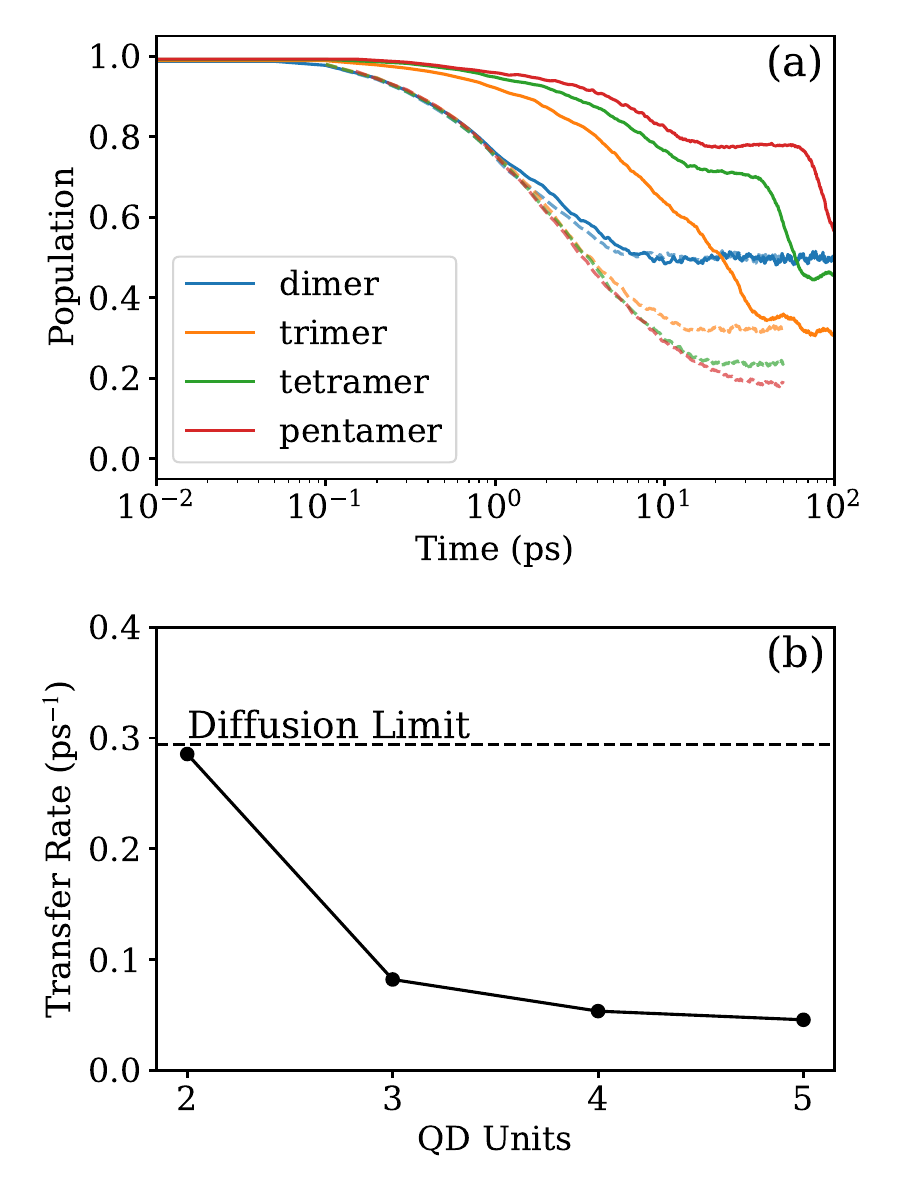} }}
    \caption{(a) Donor population dynamics in the nonadiabatic regime with [100] attachment. Solid lines represent the dynamics with all modes. Dashed lines show the dynamics with inter-dot superlattice phonons removed, representing a diffusion limit with couplings only to intra-dot phonon modes. (b) Comparison of the transfer rate between the donor and neighboring acceptor in QD arrays with different units. The rate is determined by an exponential fit of the donor population at short times.}
    \label{Fig_diff}
\end{figure}

The dynamics are completely different in the adiabatic limit, shown in Figure~\ref{Fig_pop} panels (d)-(f).  Here,  the populations exhibit underdamped dynamics with an oscillation frequency determined by the strength of the hybridization $J$ in the first few picoseconds. The overall timescale for reaching the equilibrium population only slightly increases with the chain length, while the dephasing times significantly decrease with the number of QD units across all attachments and crystal structures. Longer equilibrium times are observed in the [001] attachment due to the energy bias between adjacent QD units.\cite{hou_incoherent_2023} Energy bias is present in the [001] attachment due to the lack of reflection symmetry perpendicular to the QD array axis. Oscillations around equilibrium arise from strong coupling to superlattice phonons, with frequencies corresponding to those of the phonons (see Supplementary Figures~6 and~7 for comparison).

\textbf{Strong Couplings to Superlattice Phonons.} To rationalize the differences observed in the electron transfer dynamics for different attachments and crystal structures, we have calculated the spectral density defined as:
\begin{equation}
    S_{nm}(\omega) = \pi\sum_\alpha\omega_\alpha\lambda^\alpha_{nm}\delta(\omega-\omega_\alpha),
\end{equation}
where $\delta(\omega)$ is the Dirac delta function and $\lambda^\alpha_{nm} =  \frac{1}{2}\left(\frac{V_{nm}^\alpha}{\omega_\alpha}\right)^2$ is the reorganization energy between states $n$, $m$ with respect to mode $\alpha$. The diagonal spectral density $n=m$ characterizes the coupling strength to on-site energy $\varepsilon_n$ and off-diagonal spectral density $n\neq m$ characterizes couplings to hybridization energy $J_{nm}$. Figure~\ref{Fig_Jw} presents the diagonal spectral densities for donor ground state in WZ and ZB QD arrays with varying chain lengths. The modes with the strongest coupling are denoted as $\omega_{\rm bend}$, $\omega_{\rm syms}$, and $\omega_{\rm intra}$, with their identities influenced by the crystal symmetry and orientation of the attachment (see Supplementary Figure~2 for vibrational modes). We observe that the bending mode (denoted as $\omega_{\rm bend}$) exhibits strong coupling in the WZ [100] attachment, while the symmetric stretch mode (denoted as $\omega_{\rm syms}$) is significant in the WZ [001] attachment and to some extent in the ZB structure. Additionally, strong coupling to the set of intra-dot torsional modes labeled $\omega_{\rm intra}$, is observed across all attachments. The spectral density remains consistent regardless of the QD chain length for the torsional mode but increases significantly for the bending/symmetric stretch modes, except for the ZB structure. In the nonadiabatic limit but not in the adiabatic limit, as discussed below, this increase in coupling to the inter-dot modes causes a significant slowdown of the electron transfer dynamics, as observed for the WZ but not for the ZB attachments (see Figure~\ref{Fig_pop}). 

This slowing down of electron transfer times with longer QD chains is distinct from the slowing down caused by diffusion over longer distances. Figure~\ref{Fig_diff} panel (a) compares the donor population dynamics with (solid lines) and without (dashed lines) couplings to inter-dot superlattice phonons. Including the superlattice phonon modes results in slower electron transfer times for chain lengths greater than three QDs, compared to the diffusion limit where those modes have been removed. As evident in panel (a), the full-mode population of the donor decays to a plateau at intermediate times for longer chains, with a timescale corresponding to half the period of the inter-dot bending mode. This plateau does not appear in the diffusion limit. We extract the transfer rates shown in Figure~\ref{Fig_diff} panel (b) by fitting the population dynamics to an exponential decay. The transfer rates between the neighboring sites shown by the dashed line are independent of the chain length in the diffusion limit, while the transfer rates decrease with the number of QD units when including the inter-dot phonons.  The diffusive limit is naturally observed for the ZB structures in Figure~\ref{Fig_pop}(c) where the couplings to the inter-dot superlattice phonons are weak.


\begin{figure*}[ht!]
    \centering
    {{\includegraphics[width=17cm]{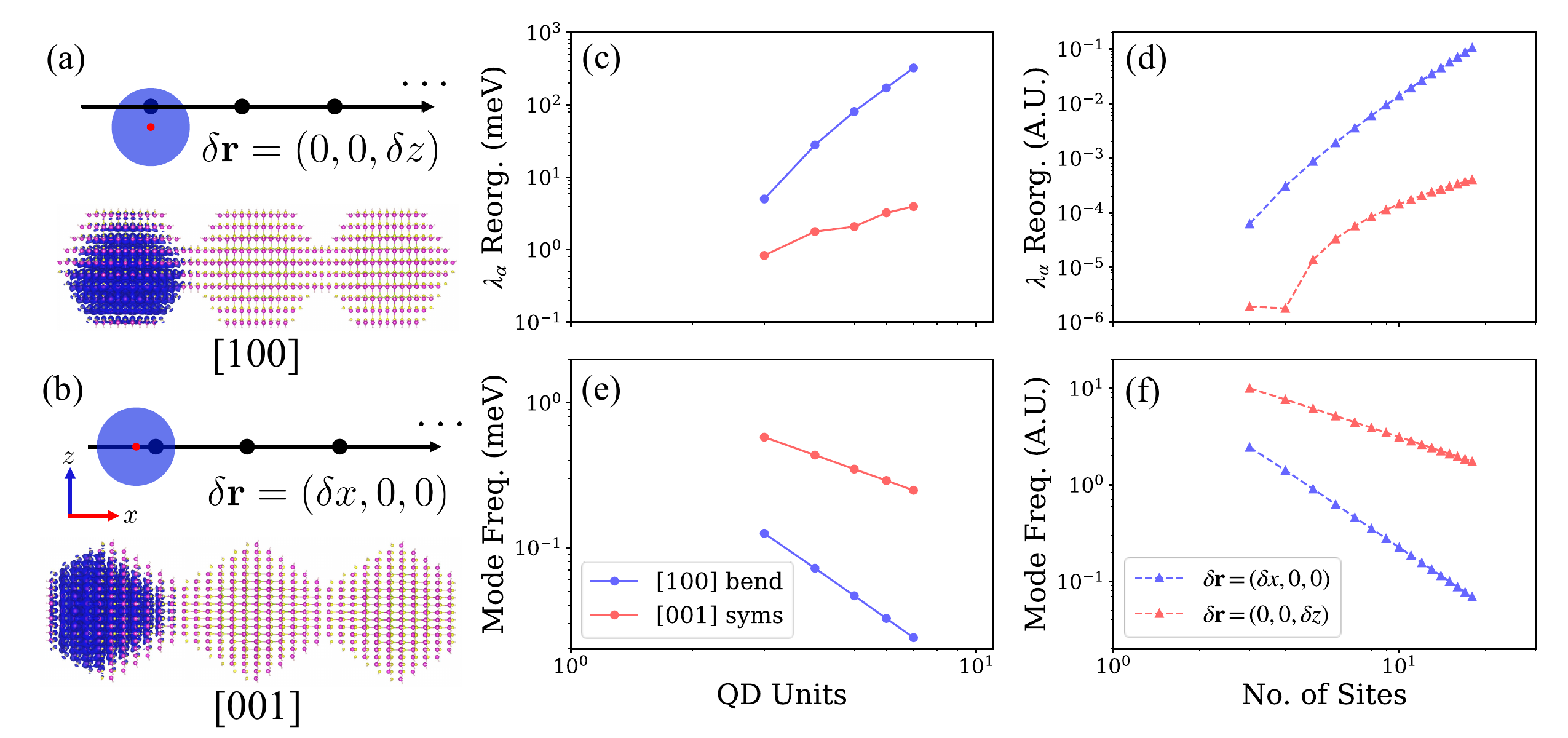} }}
    \caption{A sketch of a lattice site model chain with Gaussian-type orbitals shifted by $\delta\mathbf{r} = (0,0,\delta z)$ (panel (a)) corresponding to the WZ [100] attachment or by $\delta\mathbf{r} = (\delta x,0,0)$ (panel (b)) corresponding to the WZ [001] attachment. Panel (c) shows the reorganization energies of the most strongly coupled modes (bending and symmetric stretching) versus the number of QD units for the WZ [100] and [001] attachments. Panel (d) shows the reorganization energies of the most strongly coupled modes versus the number of sites. Panels (e) and (f) display the normal mode frequencies of the most strongly coupled modes versus the number of QD units (e) or sites (f) for the corresponding attachments in panels (c) and (d).}
    \label{Fig_model_compare}
\end{figure*}

\textbf{Understand the charge localization with atomic site model.} The coupling strength and reorganization energy of the QD superlattice phonons are closely related to the symmetry of the vibrational modes and electronic states. To gain a better understanding of the behavior of the spectral density for different attachments, we consider a simplified model on a one-dimensional lattice, schematically illustrated in Figure~\ref{Fig_model_compare}, panels (a) and (b). This enabled us to derive scaling laws for the reorganization energies and comprehend the role of symmetry breaking on the coupling to various inter-dot vibrational modes. The model comprises a single $s$-like Gaussian orbital either centered around each site or displaced by $\delta \mathbf{r}$ relative to the site. The bending and stretching vibrations were described using nearest-neighbor two- and three-body terms. A detailed description of this lattice site model is provided in the Supplementary Information section IV.

We used the displacement $\delta \mathbf{r}$ to account for the asymmetry observed in the localized electron states across different QD attachments and crystal structures, as determined from pseudopotential calculations. For the [100] attachment shown in panel (a) the electron density is shifted perpendicular to the chain axis ($\delta {\bf r} = (0,0,\delta z)$) while for [001] attachment shown in panel (b) it is shifted along the chain axis ($\delta {\bf r} = (\delta x,0,0)$). The charge density for the ZB structure (not shown) remains unshifted. We find that when the charge density is shifted perpendicular to the chain axis, the electron predominantly couples to the bending mode, consistent with the pseudopotential calculations for the WZ [100] attachment. Conversely, when the charge density is shifted along the chain axis, the electron couples more strongly to the symmetric stretch. This can be rationalized by considering the symmetry of the electronic states and the vibrational modes.

In the single-site chain model, the bending and symmetric stretch correspond to the irreducible representations $E_{1u}$ and $A_g$, respectively. For a centered orbital ($\delta {\bf r}=(0,0,0)$), the diagonal electron-phonon coupling $V_{nn}^\alpha$ vanishes for the bending mode but is nonzero for the symmetric stretch based on group theory.\cite{kouppel1984multimode} Symmetry breaking along or perpendicular to the inter-dot axis enhances the couplings to the superlattice phonons. When symmetry breaks perpendicular to the chain, the coupling to the bending mode becomes dominant and increases slightly with the chain length. Conversely, when symmetry breaks along the chain, the coupling to the bending mode vanishes and the coupling to the symmetric stretch becomes dominant as illustrated in Supplementary Figure~4.

In Figure~\ref{Fig_model_compare}, panels (c) and (d), we plot the reorganization energy of the most strongly coupled mode as a function of the number of QD units or sites. The similarity between the pseudopotential calculations (panel (c)) and the simplified model (panel (d)) at the relative range (up to hexamer) is remarkable. This correspondence provides strong validation for the relevance of the simplified site model. The reorganization energy in both calculations increases more rapidly with $N_{\rm dot}$ for the bending mode ([100] attachment) compared to the symmetric stretch ([001] attachment).  The reorganization energy is given by the ratio of the electron-phonon coupling and the vibrational frequency,  $\lambda^\alpha_{nm} =  \frac{1}{2}\left(\frac{V_{nm}^\alpha}{\omega_\alpha}\right)^2$.  The electron-phonon couplings show weak dependence on the number of QD units/sites as in Supplementary Figure~4 panel (a) and (b) and thus, the dependence of $\lambda^\alpha_{nm}$ on $N_{\rm dot}$ is induced by the behavior of $\omega_\alpha$.  Within the simplified model, these frequencies (see Supplementary Figure~3) can be expressed as (see Supplementary Information section IV for derivations)
\begin{align}
    \omega_{\rm syms} &= 2\sqrt{k}\left| \sin\left({\frac{\pi}{2N_{\rm dot}}} \right)\right|
\end{align}
and
\begin{align}
    \omega_{\rm bend} &\approx 4\sqrt{k}\sin^2\left({\frac{2\sqrt{3}\pi}{9N_{\rm dot}}}\right),
\end{align}
with an asymptotic limit given by $N_{\rm dot}^{-2}$ and $N_{\rm dot}^{-1}$ for the bending and symmetric stretch, respectively.  In Figure~\ref{Fig_model_compare}, panels (e) and (f) we plot the calculated frequencies as a function of the $N_{\rm dot}$, confirming numerically the above scaling. This implies that the reorganization energies asymptotically scale as $N_{\rm dot}^{4}$ and $N_{\rm dot}^{2}$ for the bending and symmetric stretch, respectively. 

\begin{figure*}[ht!]
    \centering
    {{\includegraphics[width=16cm]{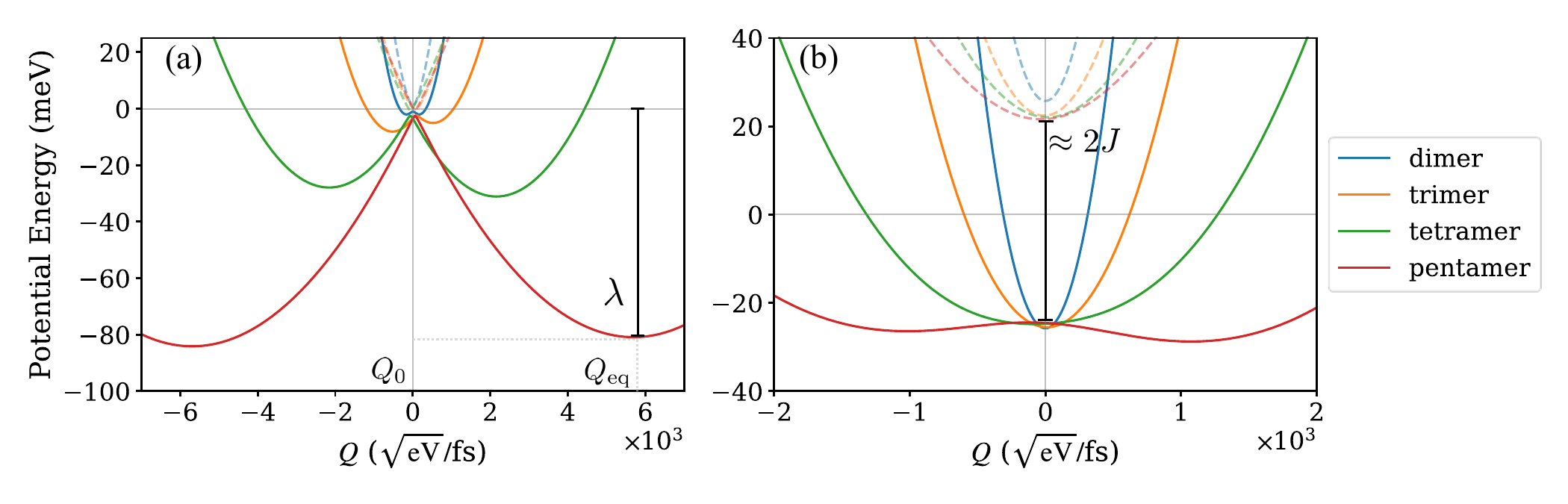} }}
    \caption{The projection of the potential energy surfaces along the primary bending mode coordinate $Q$ in the [100] WZ nonadiabatic (panel (a)) and adiabatic (panel (b)) limits for different QD chain lengths. Only the first two states on the neighboring sites are shown. The solid lines represent the ground donor potential energy surfaces, and the dashed lines represent the excited donor potential energy surfaces. $Q_0$ is the ground state reference configuration, and $Q_{\rm eq}$ is the equilibrium configuration for the donor state. Reorganization energy $\lambda$ is the energy difference between $Q_0$ and $Q_{\rm eq}$ configuration. The splitting between the two potential energy surfaces is approximately two times the electronic coupling $J$. }
    \label{Fig_pes}
\end{figure*}

\textbf{Potential energy surfaces along superlattice modes.} The increase in the reorganization energy with the length of the QD chain and the slowdown of electron transfer across the QD array in the nonadiabatic regime can be explained by considering the adiabatic potential energy surfaces along the superlattice mode that couples strongly to the excess electron, as shown in Figure~\ref{Fig_pes} panel (a). As the reorganization energy increases with the length of QD assembly, the energy barrier between the neighboring QDs increases. This results in slower electron transfer in WZ [100] structures, as shown in Figure~\ref{Fig_pop} panel (a). In the adiabatic regime, on the other hand, the energy barrier does not vary significantly with the number of QD units as shown in Figure~\ref{Fig_pes} panel (b) and the potential barrier is below the thermal energy at $T=300$ K. This is primarily because the large neck width between the QDs leads to rigid superlattice vibrations with higher phonon frequencies compared to the nonadiabatic limit. Supplementary Figures~5-7 show the effects of removing the inter-dot phonons on the transfer dynamics in both nonadiabatic and adiabatic regimes. We found that, in general, inter-dot phonons lead to faster transfer or dephasing rates when the couplings are not too strong. However, as the reorganization energies increase with the QD chain length, the electron eventually becomes localized. The system will transition to the localized regime unless $J$ is increased simultaneously to keep it in the adiabatic limit, maintaining small barriers between neighboring sites. In this case, the superlattice phonons induce faster dephasing, as shown in Figure~\ref{Fig_pop} panel (d)-(f). In general, superlattice inter-dot phonons lead to larger energy barriers in the nonadiabatic regime while facilitating dephasing in the adiabatic regime.

\section{Conclusion}
In conclusion, we have theoretically examined electron transfer in an array of interconnected colloidal QDs. The electronic structure and electron-phonon interactions were computed using semi-empirical pseudopotential methods, and the nuclear vibrations were obtained from a force field. Our results reveal that electron transfer dynamics are profoundly influenced by superlattice phonons in the nonadiabatic regime, particularly when the neck connecting neighboring QDs is narrow. Through a simplified site model, we further illustrate that the impact of coupling to superlattice phonons is related to the symmetry of the electronic wavefunctions and inter-dot vibrations.

Specifically, the lack of centrosymmetry in the wurtzite structure results in strong couplings with the inter-dot bending and symmetric stretching modes. The electron-phonon couplings and reorganization energy increase with the number of QDs in the array, thereby localizing electron transfer. In contrast, adiabatic electron transfer in QD arrays with wide neck widths is less influenced by superlattice phonons, enabling rapid transfer across multiple QD sites. While the coupling to superlattice bending modes may be suppressed in periodic 2D or 3D QD superlattices attached to a substrate, we believe that our finite-size 1D model can still elucidate the electron dynamics resulting from local vibrations across several QDs in the assembly, whether it is in 1D, 2D, or 3D. 

A few aspects neglected in our current approach, which will be the subject of future studies, include disorder resulting from size inhomogeneity of the QD monomers, the effects of ligands and solvent, as well as strain and composition. For the inhomogeneous disorder, we estimated the localization length using a 1D site chain model and found that the population dynamics may be further affected by Anderson localization with a 5\% QD size distribution in the nonadiabatic limit, where the coupling between QD sites is small. In contrast, in the adiabatic regime, the estimated Anderson localization length is an order of magnitude longer than the length scales currently studied, resulting in a negligible contribution. Additionally, the presence of surface ligands and solvent can introduce a secondary bath coupled to the electron. Strain and composition effects, such as off-stoichiometry in the QD monomers, can create defect states that influence the transfer dynamics. Future studies will incorporate an atomistic description, particularly focusing on ligand motion and various defect states, to investigate their effects on electron transfer in the QD chain.


\section{Methods}
\textbf{Semi-Empirical Pseudopotential Method.} The semi-empirical pseudopotential method was employed to calculate the quasi-electron Hamiltonian describing the valence electron of the QD arrays. The local screened pseudopotentials were fitted to a functional form in the momentum space as~\cite{fu1997local, zunger1997pseudopotential, wang1999electronic,jasrasaria_interplay_2021}
\begin{equation}\label{eq:pseudo}
\Tilde{v}(q)=a_0\left[1+a_4 \operatorname{Tr} \bm{\epsilon}\right] \frac{q^2-a_1}{a_2 \exp \left(a_3 q^2\right)-1}
\end{equation}
where $q$ is the magnitude of momentum vector, $\bm{\epsilon}$ is the strain tensor, and $a_0$ to $a_4$ are fitting parameters for Cd, Se, and S, which are tabulated in the Supplementary Table~1. The parameters were fitted simultaneously to the band structures and deformation potentials of bulk CdSe and CdS to capture the electronic and vibronic properties of the nanocrystals. Filter diagonalization technique was then applied to obtain quasiparticle electron states $\psi_i(\mathbf{r})$ of the Hamiltonian ${h}_{\rm e}(\mathbf{r})$ above the conduction band edge. The calculations were performed on real-space grids with spacings smaller than 0.7 a.u. to ensure that the eigenenergies converged to within $10^{-3}$ meV and that the sequential calculations of electron-phonon couplings were also converged.

To study electron transfer within the QD array, we constructed a set of states $\phi_n(\mathbf{r})$ localized on each QD unit, defining a set of donor and acceptor states. The transformation between the delocalized eigenstates $\psi_i(\mathbf{r})$ and the localized states $\phi_n(\mathbf{r})$ was achieved using the F\"{o}rster-Boys localization scheme, which maximizes the self-extension criterion~\cite{foster_canonical_1960}. Eigenstates with $s$- and $p$-like envelope functions were included in the localization procedure. 

\textbf{Electron-Phonon Couplings.} The first-order atomistic electron-nuclear coupling $V_{nm}^{\mu k}$ between states $\phi_n(\mathbf{r})$ and $\phi_m(\mathbf{r})$ was calculated by~\cite{jasrasaria_interplay_2021}
\begin{equation}
    V_{nm}^{\mu k} = \bra{\phi_n}\frac{\partial h_{\rm e}}{\partial R_{\mu k}}\ket{\phi_m},
\end{equation}
where $R_{\mu k}$ is the atomic coordinates with $\mu=1\dots N_{\rm atom}$ and $k=x,y,z$. In QD arrays with a narrow neck width, the superlattice vibrations have low frequencies. To avoid mixing inter-dot superlattice modes with global translational and rotational motions in this narrow neck limit, we defined a projected electron-nuclear coupling~\cite{shu_conservation_2020} to eliminate those degrees of freedom (see Supplementary Information for details). The QD array structures are relaxed using the 3-body Stillinger-Weber force field~\cite{zhou_stillinger-weber_2013}. Under the harmonic approximation, the normal mode frequencies $\omega_\alpha$ and coordinates $\mathbf{E}^\alpha$ can be calculated by diagonalizing the Hessian matrix at the equilibrium configuration. The transformation between normal mode and atomic coordinate is defined as $Q_\alpha  = \sum_{\mu k}\sqrt{m_\mu} (R_{\mu k}-R_{\mu k,0}) E_{\mu k}^\alpha$, which leads to the definition of electron-phonon couplings
\begin{equation}
    V_{nm}^\alpha = \sum_{\mu,k}\frac{1}{\sqrt{m_\mu}} E_{\mu k,\alpha}V_{nm}^{\mu k}.
\end{equation}
All the parameters are then plugged into the model Hamiltonian in Eqs.~\ref{eq:H_model} for the dynamical calculation. 

\section*{acknowledgement}
This work was supported in part by the NSF–BSF International Collaboration in the Division of Materials Research program, NSF grant number DMR-2026741 and BSF grant number 2020618. Methods used in this work were provided by the center on ``Traversing the death valley separating short and long times in non-equilibrium quantum dynamical simulations of real materials", which is funded by the U.S. Department of Energy, Office of Science, Office of Advanced Scientific Computing Research and Office of Basic Energy Sciences, Scientific Discovery through Advanced Computing (SciDAC) program, under Award No. DE-SC0022088. E. Rabani acknowledges the Alexander von Humboldt Foundation for the generous Humboldt Research Award.

\section*{Reference}

\bibliographystyle{naturemag}
\bibliography{main}

\end{document}


\title{Supplementary Information: The Role of Superlattice Phonons in Modulating Charge Transfer Across Quantum Dot Arrays}
\author{Bokang Hou$^{1*}$, Matthew Coley-O'Rourke$^1$, Uri Banin$^2$, Michael Thoss$^3$, Eran Rabani$^{1,4,5}$}
\email{Corresponding authosr. Email: bkhou@berkeley.edu, eran.rabani@berkeley.edu}
\affiliation{$^1$Department of Chemistry, University of
California, Berkeley, California 94720, United States.\\
$^2$Institute of Chemistry and the Center for Nanoscience and Nanotechnology, The Hebrew University of Jerusalem, 91904 Jerusalem, Israel.\\
$^3$Institute of Physics, University of Freiburg, Hermann-Herder-Straße 3, 79104 Freiburg, Germany.\\
$^4$Materials Sciences Division, Lawrence Berkeley National Laboratory, Berkeley, California 94720, United States.\\
$^5$The Raymond and Beverly Sackler Center of Computational Molecular and Materials Science, Tel Aviv University, Tel Aviv 69978, Israel.}

\maketitle

\section{QD Array Strucuture}
\begin{figure}[ht!]
    \centering
    \includegraphics[width=11cm]{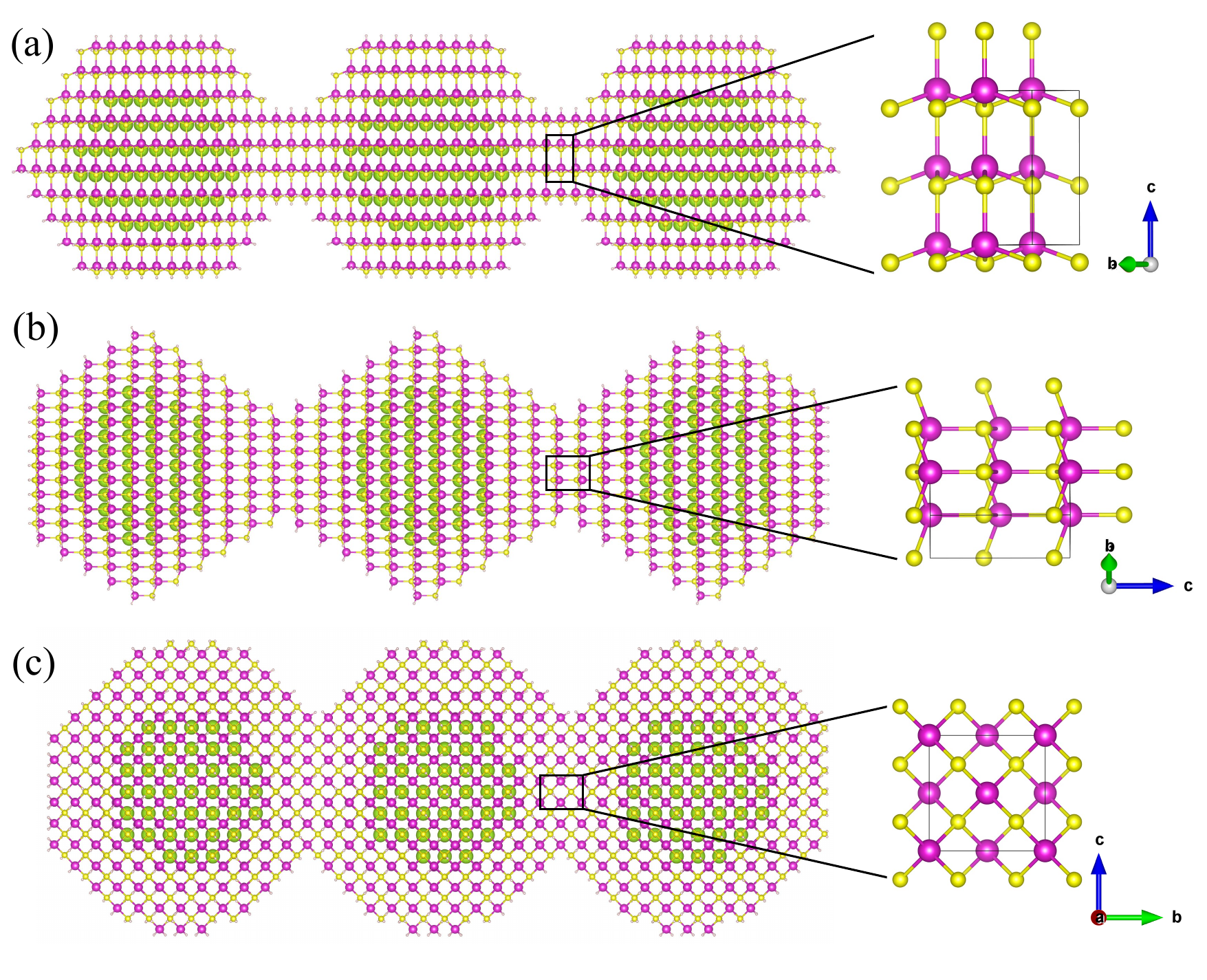}
    \caption{Structures of CdSe/CdS core/shell QD arrays (using trimer as an example) for (a) wurtzite [100] attachment (b) wurtzite [001] attachment and (c) zincblende structures. Cadmium, sulfur, and selenium atoms are depicted as magenta, yellow, and green, respectively (the size of the Se atom is enlarged for visualization). The white atoms at the surface are pseudohydrogen ligands used for passivation. The zoomed-in section shows the lattice structures, with the unit cells highlighted in boxes.}
    \label{fig:qd_arr_struct}
\end{figure}
The CdSe/CdS core-shell colloidal quantum dot (QD) was constructed by adding CdS shells to a faceted CdSe core seed with a lattice constant of bulk wurtzite CdSe ($a=4.3$ \AA, $c=\sqrt{\frac{8}{3}}a$). Each QD unit has a diameter of 3.5 nm with a CdSe core of 2.0 nm. The wurtzite QD arrays were constructed by attaching multiple QDs through the $[100]$ ([10$\bar{1}$0]) or $[001]$ ([0001]) crystalline planes as shown in Supplementary Figure~\ref{fig:qd_arr_struct}. To simplify the discussion and unify the coordinate system for all structures, we define the $c$-axis to coincide with the $z$-axis in the [100] orientation. Note that [100] orientation results in a mirror plane perpendicular to the attachment axis, while [001] orientation does not. The neck of the dimer can be widened by attaching CdS mono-layers to the connection area between two QDs. The structures were minimized with Stillinger-Weber force field parameterized for II-VI nanostructures implemented in LAMMPS~\cite{plimpton_fast_1995,zhou_stillinger-weber_2013}.

\section{Electronic and Vibronic Structure Calculations of QD Array }
\textbf{Semi-Empirical Pseudopotential Method:} The semi-empirical pseudopotential model was employed to calculate the quasi-electron Hamiltonian describing the valence electron of the QD arrays. The local screened pseudopotentials were fitted to a functional form in the momentum space as 
\begin{equation}\label{eq:pseudo}
\Tilde{v}(q)=a_0\left[1+a_4 \operatorname{Tr} \bm{\epsilon}\right] \frac{q^2-a_1}{a_2 \exp \left(a_3 q^2\right)-1}
\end{equation}
where $q$ is the magnitude of momentum vector, $\bm{\epsilon}$ is the strain tensor, and $a_0$ to $a_4$ are fitting parameters for Cd, Se, and S, which are tabulated in the Supplementary Table~1. The parameters were fitted simultaneously to the bulk band structures and deformation potentials of CdSe and CdS to capture the electronic and vibronic properties of the nanocrystals. The real-space quasi-electron Hamiltonian ${h}_{\rm e}(\mathbf{r})$ was given by 
\begin{equation}
    {h}_{\rm e}(\mathbf{r}) = -\frac{1}{2}\nabla^2_{\mathbf{r}} + \sum_\mu v_\mu\left(|\mathbf{r}-\mathbf{R}_{0,\mu}|\right),
\end{equation}
where $v_\mu(\mathbf{r})$ is the real-space pseudopotential for atom $\mu$, obtained from the Fourier transform of $\Tilde{v}(q)$. The filter diagonalization technique was then applied to obtain quasiparticle electron states $\psi_i(\mathbf{r})$ above the conduction band edge. The calculations were performed on real-space grids smaller than 0.7 a.u., ensuring eigenenergies converged to less than $10^{-3}$ meV, and sequential calculations of electron-phonon couplings were converged.

\renewcommand{\thetable}{1}
\begin{table}[ht]
\begin{tabular}{|c|c|c|c|c|c|}
\hline
   & $a_0$    & $a_1$       & $a_2$     & $a_3$      & $a_4$     \\ \hline
Cd & -31.4518 & 1.3890 & -0.0502 & 1.6603 & 0.0586 \\ \hline
Se & 8.4921 & 4.3513   & 1.3600 & 0.3227  & 0.1746 \\ \hline
S  & 7.6697 & 4.5192   & 1.3456 & 0.3035  & 0.2087 \\ \hline
\end{tabular}
\caption{Pseudopotential parameters in Eq.~\ref{eq:pseudo} for Cd, Se, and S atoms. All parameters are given in atomic units.}
\label{tab:3}
\end{table}

The quasi-electron eigenstates $\psi_i(\mathbf{r})$ are delocalized across the QD array with $N_{\rm dot}$ units. To study electron transfer within the QD array, we constructed a set of states $\phi_n(\mathbf{r})$ localized on each QD unit, defining a set of donor and acceptor states. The transformation between the delocalized eigenstates $\psi_i(\mathbf{r})$ and the localized states $\phi_n(\mathbf{r})$ was achieved using the F\"{o}rster-Boys localization scheme, which maximizes the self-extension criterion~\cite{foster_canonical_1960}
\begin{equation}
    \left< \hat{\Omega}\right>_{FB} = \sum_{n\in\mathcal{D},\mathcal{A}}\left(\int \mathrm{d}^3\mathbf{r} |\phi_n(\mathbf{r})|^2\mathbf{r}\right)^2.
\end{equation}
Expectation $\left< \hat{\Omega} \right>_{FB}$ was maximized through successive $2\times 2$ Jacobi rotations of wavefunction pairs. The resulting localized states are related to the eigenstates by a unitary transformation $U$
\begin{equation}
    \phi_n(\mathbf{r}) = \sum_{i}U_{ni}\psi_i(\mathbf{r}). 
\end{equation}
Eigenstates with $s$- and $p$-like envelope functions were included in the localization procedure. The resulting localized states comprise a set of $s$ and $p$ quasi-electron states on each QD unit.

\textbf{Vibronic Coupling:} The first-order atomistic electron-nuclear coupling $V_{nm}^{\mu k}$ between states $\phi_n(\mathbf{r})$ and $\phi_m(\mathbf{r})$ was calculated by~\cite{jasrasaria_interplay_2021}
\begin{equation}
    V_{nm}^{\mu k} = \bra{\phi_n}\frac{\partial h_{\rm e}}{\partial R_{\mu k}}\ket{\phi_m} = \bra{\phi_n}\frac{\partial v_{\mu}\left(|\mathbf{r}-\mathbf{R}_{0,\mu}|\right)}{\partial R_{\mu k}}\ket{\phi_m}\equiv \int d\mathbf{r}{\phi_n}^*(\mathbf{r})\frac{\partial v_{\mu}\left(|\mathbf{r}-\mathbf{R}_{0,\mu}|\right)}{\partial R_{\mu k}}\phi_m(\mathbf{r}),
\end{equation}
where $R_{\mu k}$ is the atomic coordinates with $\mu=1\dots N_{\rm atom}$ and $k=x,y,z$. In QD arrays with a narrow neck width, the super-lattice vibrations have low frequencies. To avoid mixing inter-dot super-lattice modes with global translational and rotational motions in this narrow neck limit, we defined a projected electron-nuclear coupling $\Bar{V}_{nm}^{\mu k}$ to eliminate these global motions
\begin{equation}
    \bar{V}_{nm}^{\mu k} = \sum_{\mu' k'}({\delta_{\mu k, \mu' k'} - \mathcal{Q}_{\mu k, \mu' k'})V_{nm}^{\mu' k'}}.
\end{equation}
The projector $\mathcal{Q}$ is defined as~\cite{shu_conservation_2020}
\begin{equation}
    \mathcal{Q}_{\mu k, \mu' k'} = \frac{1}{N}\delta_{k,k'} + \sum_{i,j,i',j'}\Tilde{\epsilon}_{ijk} R_{\mu i} [\mathbf{\Tilde{I}}^{-1}]_{jj'}\Tilde{\epsilon}_{i'j'k'}R_{\mu' i'}
\end{equation}
where $i,j,k,i',j',k'$ label $x,y,z$ directions and $\mu, \mu'$ label atoms. $\mathbf{\Tilde{I}}^{-1}$ is the inverse of the reduced moment of inertia with all masses set to 1. $\bm{\Tilde{\epsilon}}$ is a third-order unit pseudotensor related to the Levi-Civita symbol. This projector removed the center of mass translation and rotation from the original coupling. The projected coupling satisfies $\sum_{\mu k} \bar{V}_{nm}^{\mu k} \equiv 0$ for given $n, m$.

The QD array structures are relaxed using the 3-body Stillinger-Weber force field. Under the harmonic approximation, the normal mode frequencies $\omega_\alpha$ and coordinates $\mathbf{E}^\alpha$ can be calculated by diagonalizing the Hessian matrix at the equilibrium
\begin{equation}
    D_{\mu k,\mu' k'} = \frac{1}{\sqrt{m_\mu m_{\mu'}}}\left.\frac{\partial^{2} U_\text{SW}}{\partial u_{\mu k}\partial u_{\mu' k'}}\right|_{u=0},
\end{equation}
where $m_\mu$ is the mass of $\mu$-th atom and $u_{\mu k}=R_{\mu k}-R_{0,\mu k}$ is the atomic displacement from the equilibrium. The transformation between normal mode and atomic coordinate is $Q_\alpha  = \sum_{\mu k}\sqrt{m_\mu} (R_{\mu k}-R_{\mu k,0}) E_{\mu k}^\alpha$, which leads to the definition of electron-phonon couplings
\begin{equation}
    V_{nm}^\alpha = \sum_{\mu,k}\frac{1}{\sqrt{m_\mu}} E_{\mu k,\alpha}V_{nm}^{\mu k}.
\end{equation}

\begin{figure}[ht!]
    \centering
    \includegraphics[width=17cm]{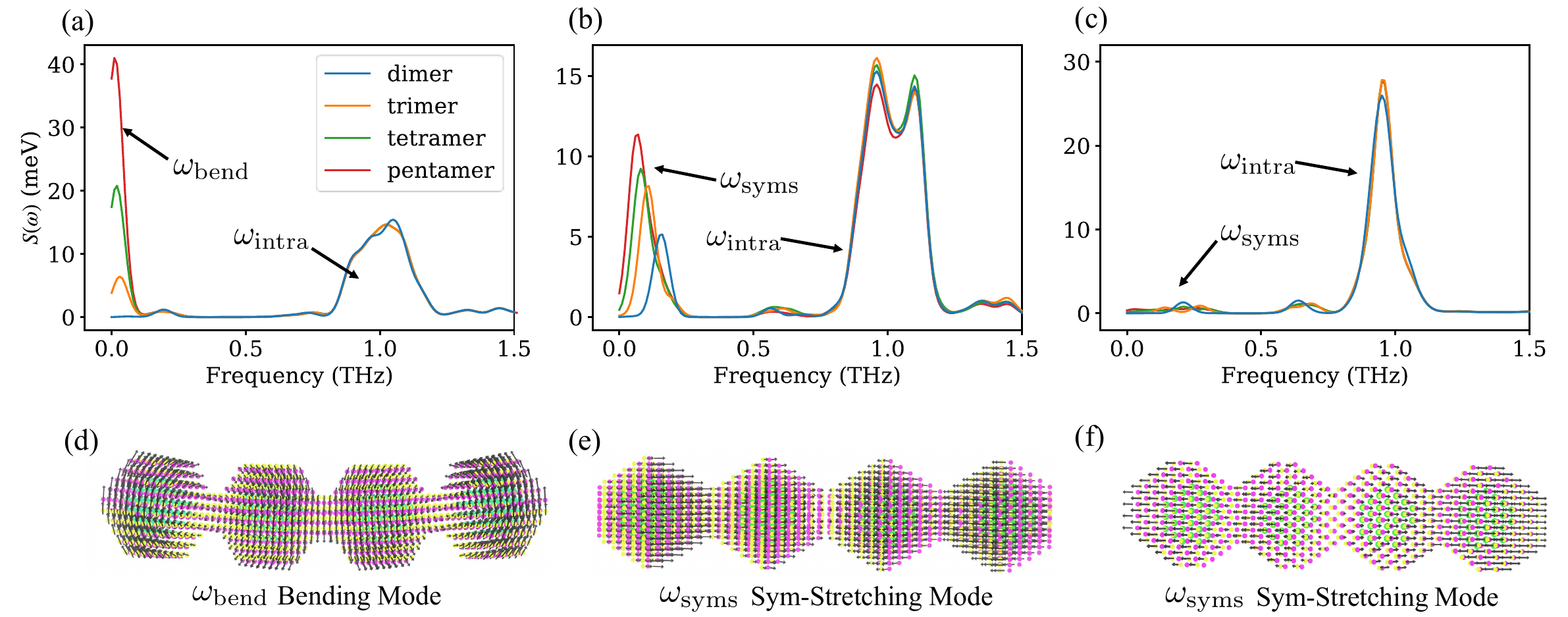}
    \caption{Zoom-in diagonal spectral density of (a) WZ [100], (b) WZ [001] and (c) ZB structures for the donor state in the nonadiabatic regimes. Superlattice phonons that significantly contribute to the reorganization energy and dynamics are shown in (d-f).}
    \label{fig:Fig_SI_Jw}
\end{figure}

\section{Spin Mapping Dynamics}
As discussed in the main text, the model Hamiltonian $H_{\rm QD}$ for QD arrays includes three parts that can be written more compactly as 
\begin{align}
\begin{split}
    H_{\rm QD} &= H_{\rm nu} + H_{\rm el} + H_{\rm el-ph}\\
     &= \sum_{\alpha}^{N_{\rm vib}}\frac{1}{2}{P}_{\alpha}^2 + \frac{1}{2}\omega_\alpha^2Q_\alpha^2 + \sum_{n,m}^{N_{\rm el}} \left(\varepsilon_{n}\delta_{nm} + J_{nm} + \sum_\alpha^{N_{\rm vib}} V_{nm}^\alpha Q_\alpha\right) \ket{\phi_n}\bra{\phi_m}
\end{split}
\end{align}
where $\delta_{nm}$ is the Kronecker Delta function, and other quantities are defined in the main text. The general idea of spin mapping is to map the quantum mechanical operators $\ket{\phi_n}\bra{\phi_m}$ in $H_{\rm QD}$ to a set of classical mapping variables $\left\{p_n, q_n\right\}$. Following the procedure described in Ref.~\cite{runeson_generalized_2020}, we chose the following mapping scheme
\begin{align}
    \ket{\phi_n}\bra{\phi_m} \mapsto \frac{1}{2\hbar} \left({q}_n{q}_m+{p}_n {p}_m-\hbar\gamma_s \delta_{n m}\right),
\end{align}
where the zero point energy parameter is related to system dimension $N_{\rm el}$ by $\gamma_s = 2(\sqrt{N_{\rm el}+1}-1)/N_{\rm el}$. This chose of $\gamma_s$ is also called the Wigner representation in the Stratonovich–Weyl transformation. The mapped Hamiltonian $\mathcal{H}_{\rm QD}$ now takes the form of 
\begin{equation}
    \mathcal{H}_{\rm QD} = \sum_{\alpha}^{N_{\rm vib}}\frac{1}{2}{P}_{\alpha}^2 + \frac{1}{2}\omega_\alpha^2Q_\alpha^2 + \frac{1}{2\hbar}\sum_{n,m}^{N_{\rm el}} \left(\varepsilon_{n}\delta_{nm} + J_{nm} +  \sum_\alpha^{N_{\rm vib}} V_{nm}^\alpha Q_\alpha\right)  \left({q}_n{q}_m+{p}_n {p}_m-\hbar\gamma_s \delta_{n m}\right)
\end{equation}
where $\left\{p_n, q_n\right\}$ and $\left\{P_\alpha, Q_\alpha\right\}$ are two sets of classical phase space variables describing electronic and nuclear degrees of freedom respectively. The equations of motion are 
\begin{equation}
\begin{aligned}\label{eq:eom}
\dot{q}_n & = \frac{\partial \mathcal{H}_{\rm QD}}{\partial p_n} = \frac{1}{\hbar}\sum_m^{N_{\rm el}} (\varepsilon_n\delta_{nm}+J_{nm}+\sum_\alpha V_{nm}^\alpha Q_\alpha) p_m, \\
\dot{p}_n & = -\frac{\partial \mathcal{H}_{\rm QD}}{\partial q_n} = -\frac{1}{\hbar}\sum_m^{N_{\rm el}} (\varepsilon_n\delta_{nm}+J_{nm}+\sum_\alpha V_{nm}^\alpha Q_\alpha) q_m, \\
\dot{Q}_\alpha & = \frac{\partial \mathcal{H}_{\rm QD}}{\partial P_\alpha} = {P}_\alpha, \\
\dot{P}_\alpha & = -\frac{\partial \mathcal{H}_{\rm QD}}{\partial Q_\alpha} =-\frac{\partial U}{\partial Q_\alpha}-\frac{1}{2\hbar}\sum_{n, m>n}^{N_{\rm el}} V_{nm}^\alpha\left(q_n q_m+p_n p_m-\hbar\gamma_s\delta_{n m}\right) .
\end{aligned}
\end{equation}
For the initial condition, the mapping variables $\{p_n, q_n\}$ were sampled uniformly from a unit circle in terms of the action-angle representation
\begin{align}
    (q_n,p_n)\mapsto(I_n, \varphi_n), \hspace{0.5cm} q_n + ip_n = \sqrt{2I_n+\hbar\gamma_s}e^{-i\varphi_n}
\end{align}
where $\varphi_n\in[0,2\pi)$.  The harmonic nuclear degrees of freedom were sampled from the Boltzmann distribution
\begin{align}
    Q_\alpha\sim\mathcal{N}\left(0,\frac{\sqrt{k_BT}}{\omega_\alpha}\right), \hspace{0.5cm}P_\alpha\sim\mathcal{N}\left(0,\sqrt{{k_BT}}\right)
\end{align}
where $\mathcal{N}(\mu, \sigma)$ is the normal distribution. Integration of the equations of motion in Eqs.~\ref{eq:eom} was done by the split-Liouvillian integrator~\cite{richardson_analysis_2017}. The population for each initial condition can be retrieved from $\mathcal{P}_n(t) = \frac{1}{2\hbar}(p_n^2+q_n^2-\hbar\gamma_s)$. Spin-mapping approximates the time-dependent system population in the density matrix $\rho_{nn}^S(t)$ as an ensemble average of those tractories, i.e. $\rho_{nn}^S(t) \approx\left<\mathcal{P}_n(t)\right>_{\rm cl}$. We found the population converged after averaging over $10^4$ trajectories. 

\section{Lattice Site Model}
\textbf{Electronic part:} We employed a tight-binding lattice site model to analyze the scaling of electron-phonon interactions in QD units across various attachments and lattice structures. Each site represents the QD's center of mass, with a spherical Gaussian-type orbital representing the envelope function of the QD's lowest unoccupied molecular orbital (LUMO). The single-electron Hamiltonian $H_{\rm el}$ of the $N$ lattice (QD) sites can be written as
\begin{equation}
    H_{\rm el} = -\frac{1}{2}\nabla^2 - \sum_{\mu=1}^N v_\mu(|\mathbf{r}-\mathbf{R}_{\mu}|),
\end{equation}
where $\mathbf{r}$ is the electron coordinate, and $v_\mu(|\mathbf{r}-\mathbf{R}_{\mu}|)$ represents the local potential around QD site $\mathbf{R}_{\mu}$. We will take this to be a bare Coulomb potential for simplicity, i.e. $v_\mu(|\mathbf{r}-\mathbf{R}_{\mu}|) = 1/|\mathbf{r}-\mathbf{R}_{\mu}|$. The 1S atomic Gaussian orbital at site $n$ is 
\begin{align}
        \phi_n(\mathbf{r}) &= \left(\frac{2a}{\pi}\right)^{\frac{3}{4}} e^{-a|\mathbf{r}-\mathbf{R}_n^c|^2},
\end{align}
where $\mathbf{R}_n^c$ is the center of the orbital and $a$ measures the orbital radius. We use Greek letters $\mu, \nu, \dots$ to label lattice sites and Latin letters $n, m, \dots$ to label orbitals. To study how the symmetry of the wavefunction affects electron-nuclear couplings in lattice sites, we can introduce a symmetry-breaking displacement $\delta\mathbf{r}$, such that $\mathbf{R}_\mu^c = \mathbf{R}_\mu+\delta\mathbf{r}$ and the new wavefunction becomes $\phi_\mu(\mathbf{r}-\delta\mathbf{r})$. Specifically, we focus on symmetry breaking in two directions as shown in Fig.~4 (a) and (b) in the main text: $\delta\mathbf{r} = (0,0,\delta z)$, representing wavefunction shifts in the QD array attached from the [100] facet, and $\delta\mathbf{r} = (\delta x,0,0)$, representing wavefunction shifts in the QD array attached from the [001] facet.

\begin{figure}[ht!]
    \centering
    \includegraphics[width=15cm]{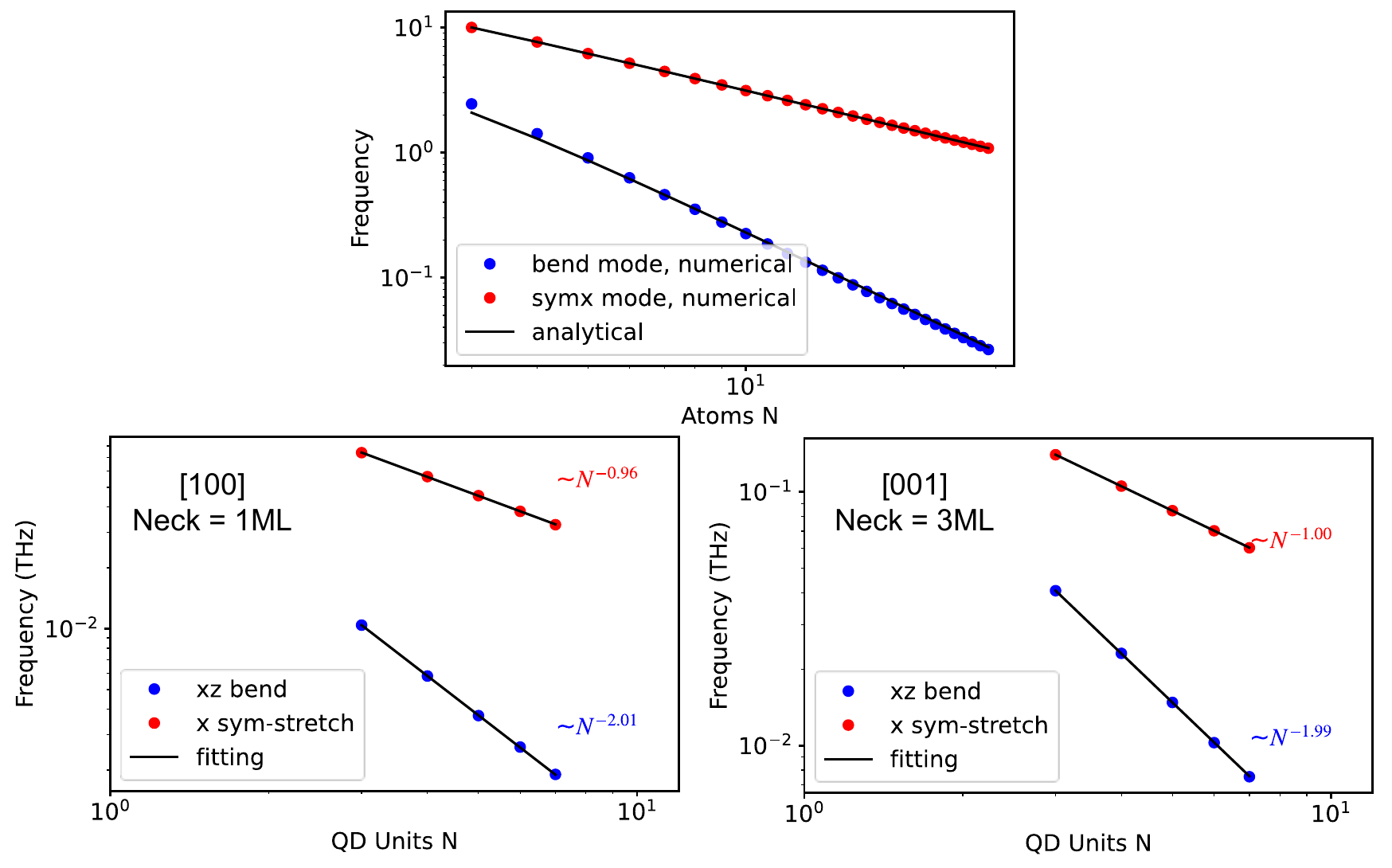}
    \caption{Frequency scaling for bending and stretching mode of the lattice sites (top) and quantum dot model (bottom). In the lattice sites model, the force constant is chosen to be $k_{\parallel} = 100k_{\perp}$ as the shear modulus is usually small.}
    \label{fig:freq_scale_bend_symx_compare}
\end{figure}

\textbf{Vibratoinal part:} The vibrations of the QDs or lattice sites were modeled as a spring-mass model. To describe both stretching and bending-type motion, we need to write the Hamiltonian in both longitudinal ($\parallel$) and transverse ($\perp$) directions
\begin{align}
    H_{\parallel} &= \sum_{i=1}^N\frac{p_{x,i}^2}{2} + \sum_{i=1}^{N-1} \frac{1}{2}k_{\parallel}(x_{i+1}-x_i)^2,\\
    H_{\perp} &= \sum_{i=1}^N\frac{p_{z,i}^2}{2} + \sum_{i=1}^{N-2} \frac{1}{2}k_{\perp}d^2\left(\frac{z_i-z_{i+1}}{d} + \frac{z_{i+2}-z_{i+1}}{d}\right)^2\\
    &= \sum_{i=1}^N\frac{p_{z,i}^2}{2} + \sum_{i=1}^{N-2} \frac{1}{2}k_{\perp}\left(z_i - 2z_{i+1} + z_{i+2}\right)^2,
\end{align}
where in the transverse direction we considered the angle change between three atoms, using the fact for small angle vibration, $\delta \theta_{i+1} \approx \tan\theta_{i,i+1} + \tan\theta_{i+1,i+2}$. The Hessian matrices in both directions are
\begin{align}
    D^{\parallel}_{ij} &= k_{\parallel}\left(-\delta_{i,j-1} + 2\delta_{i,j} - \delta_{i,j+1}\right)\\
    D^{\perp}_{ij} &= k_{\perp}\left(\delta_{i,j-2} - 4 \delta_{i,j-1} + 6\delta_{i,j} - 4 \delta_{i,j+1}+\delta_{i,j+2}\right). 
\end{align}
The longitudinal Hessian $\mathbf{D}_{\parallel}$ is tridiagonal, and can be diagonalized analytically with the normal mode frequency
\begin{align}
    \omega_\alpha^{\parallel} &= 2\sqrt{k}\left| \sin{\frac{(\alpha-1)\pi}{2N}} \right|, \alpha = 1\dots N
\end{align}
and eigenvectors $\mathbf{E}_\alpha^\parallel = (\dots E_{\mu x}^\alpha,0,0\dots)^\mathsf{T}$.
The transverse Hessian $\mathbf{D}_{\perp}$ is pentadiagonal, and the mode frequency is approximately
\begin{align}
    \omega_\alpha^{\perp} &\approx 4\sqrt{k}\sin^2{\frac{2(\alpha-1)\pi}{3\sqrt{3}N}}, \alpha = 1\dots N,
\end{align}
with the normal modes $\mathbf{E}_\alpha^\perp = (\dots 0,0,E_{\mu z}^\alpha\dots)^\mathsf{T}$. $\alpha=1$ corresponds to the transnational motion, and $\alpha=2$ corresponds to symmetric stretch and bending respectively, which are shown in the main text
\begin{align}
    \omega_{\rm syms} &= 2\sqrt{k}\left| \sin\left({\frac{\pi}{2N_{\rm dot}}} \right)\right|\\
    \omega_{\rm bend} &\approx 4\sqrt{k}\sin^2\left({\frac{2\sqrt{3}\pi}{9N_{\rm dot}}}\right). 
\end{align}


\textbf{Symmetry and vibronic couplings:} The electron-phonon coupling, or vibronic coupling, is defined as the matrix element of the first-order change in the electronic Hamiltonian with respect to the phonon coordinate. The relation between coupling to the lattice site and coupling to the nuclear vibration is given by the normal mode transformation 
\begin{align}
    {V}_{\alpha}^{nm} &= \sum_{\mu, k}E_{\mu k}^{\alpha}  V^{nm}_{\mu k},\hspace{0.5 cm} k=x, y, z
\end{align}
where $E_{\mu k}^{\alpha}$ is the coefficient for normal mode $\alpha$ and $V^{nm}_{\mu k}$ is the local site coupling defined as 
\begin{equation}
    V^{nm}_{\mu k} = \int d\mathbf{r}\phi^{*}_{n}(\mathbf{r}-\mathbf{R}_n^c) \frac{\partial H_{\rm el}}{\partial {R}_{\mu k}}\phi_{m}(\mathbf{r}-\mathbf{R}_m^c).
\end{equation}
Now we will look at the couplings with symmetry breaking in two directions: transverse ($z$) and longitudinal ($x$). 

\begin{itemize}
    \item \textbf{Transverse ($z$) Direction:}  When there is no symmetry breaking $\delta z=0$ ($\mathbf{R}^c_n = \mathbf{R}_n$), the site couplings are
    \begin{align*}
        V^{nm}_{\mu z} &= \int d\mathbf{r} \phi^{*}_{n}(\mathbf{r}-\mathbf{R}_n) \frac{\partial H_{\rm el}}{\partial {R}_{\mu z}}\phi_{m}(\mathbf{r}-\mathbf{R}_m)\\
        &= \int d\mathbf{r}\phi^{*}_{n}(\mathbf{r}-\mathbf{R}_n) \frac{r_z}{|\mathbf{r}-\mathbf{R}_\mu|^3} \phi_{m}(\mathbf{r}-\mathbf{R}_m), \hspace{0.3cm} \text{where } R_{\mu z}=0\\
        &= 0
    \end{align*}
    So in consequence, ${V}_{\mathrm{bend}}^{nm}=0$. When $\delta \mathbf{r}=(0,0,\delta z)$, the local coupling $ V^{nm}_{\mu z}\neq 0$. In particular, the diagonal elements:
    \begin{align*}
        V^{nn}_{\mu z} &= \int d\mathbf{r}  \frac{\partial H_{\rm el}}{\partial {R}_{\mu z}}|\phi_n(\mathbf{r}-\mathbf{R}_n-\delta \mathbf{r})|^2\\
         &= \left(\frac{2a}{\pi}\right)^\frac{3}{2} \int d\mathbf{r}  \frac{r_z}{[(r_x-R_{\mu z})^2 + r_y^2 + r_z^2]^\frac{3}{2}}e^{-2a[(r_x-R_{nx})^2 + r_y^2 +(r_z-\delta z)^2]}
    \end{align*}
    let $c=R_{nx}-R_{\mu x}$
    \begin{align*}
        &= \left(\frac{2a}{\pi}\right)^\frac{3}{2} \int d\mathbf{r}  \frac{r_z}{[r_x^2 + r_y^2 + r_z^2]^\frac{3}{2}}e^{-2a[(r_x-c)^2 + r_y^2 +(r_z-\delta z)^2]}
    \end{align*}
    One can see that for $c\rightarrow-c$, we have $V^{nn}_{\mu z}\rightarrow V^{nn}_{\mu z}$. So the coupling constant has even symmetry along $x$ direction. It's a good approximation to assume in the $z$-direction, $V_{\mu z}^{nm} = V_z\delta_{\mu n}\delta_{nm}$ (this is in the tight-binding limit), where $n,\mu=1\dots N$, and the constant $V_z$ should be proportional to the magnitude of asymmetric factor $\delta z$. Now the bending coupling 
    \begin{align*}
    \begin{split}
        V_{\rm bend}^{nn}  &\approx \sum_{\mu}E_{\mu z}^{\mathrm{bend}}  V^{nn}_{\mu z} \hspace{0.2cm}\text{(loc)}\\
        &= \sum_{\mu}E_{\mu z}^{\mathrm{bend}}  V_{z}\delta_{\mu n} = V_{z}  E_{n z}^{\mathrm{bend}},
    \end{split}
    \end{align*}
    which means couplings to site $n$ is closely related to the normal mode element $E_{n z}$. 
    \item \textbf{Longitudinal ($x$) Direction:} For the stretching mode along atomic axis $x$, the local coupling $V_{\mu x}^{nm}$ and eigenstate couplings $V_{\rm symx}^{ab}$ are in general nonzero regardless of $\delta\mathbf{r}$. When there is no symmetry breaking along $x$ , i.e. $\delta x=0$, $\mathbf{R}_{\mu} = (R_{\mu x}, 0, 0)$
    \begin{align*}
         V^{nn}_{\mu x} &= \left(\frac{2a}{\pi}\right)^\frac{3}{2} \int d\mathbf{r} \frac{r_x-R_{\mu x}}{|\mathbf{r}-\mathbf{R}_\mu|^3}e^{-2a|\mathbf{r}-\mathbf{R_n}|^2}\\
         &= \left(\frac{2a}{\pi}\right)^\frac{3}{2} \int d\mathbf{r} \frac{r_x-R_{\mu x}}{[(r_x-R_{\mu x})^2 + r_y^2 + r_z^2]^\frac{3}{2}}e^{-2a[(r_x-R_{n x})^2 + r_y^2 + r_z^2]}\\
         &= \left(\frac{2a}{\pi}\right)^\frac{3}{2} \int d\mathbf{r} \frac{r_x}{[r_x^2 + r_y^2 + r_z^2]^\frac{3}{2}}e^{-2a[r_x^2 + r_y^2 + r_z^2]}e^{4a c r_x}e^{-2a c^2}
    \end{align*}
    where again $c=R_{nx}-R_{\mu x}$, and for $c\rightarrow-c$, we have $V^{nn}_{\mu x}\rightarrow-V^{nn}_{\mu x}$. If we take into account of this anti-symmetry structure of the coupling constant $V_{\mu x}^{nn}$, it's a good assumption to assume that $V_{\mu x}^{nn} = V_x(\delta_{\mu,n-1}-\delta_{\mu,n+1})$, where $n=2\dots N-1$ and $\mu=1\dots N$. Then the symmetric stretch along $x$ can be written as
    \begin{align}
    {V}_{\mathrm{symx}}^{nn} &= \sum_{\mu=1}^NE_{\mu x}^{\mathrm{symx}}  V^{nn}_{\mu x}, \hspace{0.2cm}\text{(loc)}\\
    &\approx \sum_{\mu=1}^NE_{\mu x}^{\mathrm{symx}}  V_x(\delta_{\mu,n-1}-\delta_{\mu,n+1}) = V_x(E^{\rm symx}_{n-1,x}-E^{\rm symx}_{n+1,x}), 1<n<N
    \end{align}
    This means that couplings to the site $n$ along $x$ is closely related to the neighboring normal mode elements. 
\end{itemize}

\begin{figure}[ht!]
    \centering
    \includegraphics[width=16cm]{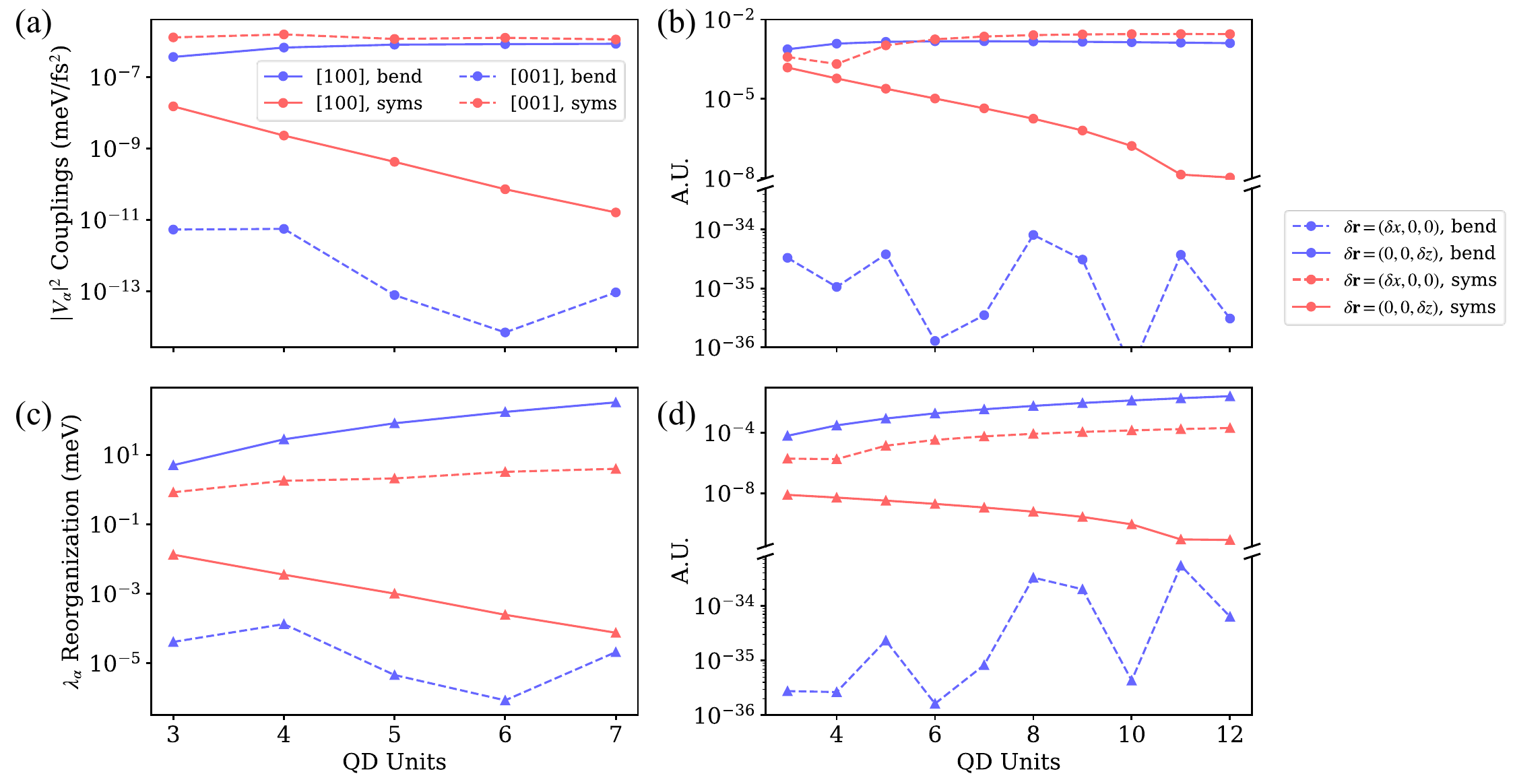}
    \caption{Comparison of the electron-phonon coupling squared $|V^\alpha|^2$ and reorganization energies $\lambda_\alpha$ in the QD arrays (a, c) and atomic site model (b, d) for bending and symmetric stretching modes.}
    \label{fig:Fig_SI_model_compare}
\end{figure}

Supplementary Figure~\ref{fig:Fig_SI_model_compare} compares the electron-phonon couplings and reorganization energies from pseudopotential calculation (panel (a) and (c)) with the QD site model discussed above (panel (b) and (d)). As shown in Supplementary Figure~\ref{fig:Fig_SI_model_compare} panel (d), When symmetry breaks along the $z$ direction (perpendicular to the chain), the bending mode couples strongly to the electronic state. Conversely, when symmetry is preserved, the bending mode does not couple. Similarly, when the symmetry breaks along the $x$ direction, the coupling to the symmetric stretch is dominant while the coupling to the bending mode vanishes. In general, the electron-phonon couplings have weaker dependence on the number of QD sites than reorganization energy.

In summary, we worked with the following Holstein-like Hamiltonian in the local electronic and atomic basis
\begin{align}
    \begin{aligned}
        H_{\rm el} &= \sum_n^{N-1} J(\ket{\phi_n}\bra{\phi_{n+1}} + h.c.)\\
        H_{\rm nu} &= H_{\parallel} + H_{\perp} = \sum_{\mu=1}^N\frac{p_{x,\mu}^2}{2} + \frac{p_{z,\mu}^2}{2} + \sum_{\mu=1}^{N-1} \frac{1}{2}k_{\parallel}(x_{\mu+1}-x_\mu)^2 + \sum_{\mu=1}^{N-2} \frac{1}{2}k_{\perp}\left(z_\mu - 2z_{\mu+1} + z_{\mu+2}\right)^2\\
        H_{\rm el-nu} &= \sum_{n=1}^N\ket{\phi_n}\bra{\phi_n}\sum_{\mu=1}^N (V_{\mu x}^n x_\mu+V_{\mu z}^n z_\mu)
    \end{aligned}
\end{align}
where $n$, $\mu$ label electronic basis and atoms respectively. The atomic electron-nuclear coupling $V_{nk}$ ($k=x,z$) is given by
\begin{equation}
    V_{\mu k}^n = \int d\mathbf{r} \phi_n^*(\mathbf{r}-\mathbf{R}_n^c)\frac{\partial H_{\rm el}}{\partial R_{\mu k}} \phi_n(\mathbf{r}-\mathbf{R}_n^c).
\end{equation}
In the normal mode representation, the longitudinal and transverse modes are given by
\begin{align}
    q_\alpha^{\parallel} = \sum_{\mu=1}^{N}E_{\mu x}^\alpha x_\mu, \hspace{0.5cm} q_\alpha^{\perp} = \sum_{\mu=1}^{N} E_{\mu z}^\alpha z_\mu,
\end{align}
where $E_{\mu x}$ ($E_{\mu z}$) is the normal mode eigenvector in longitudinal (transverse) direction. Substitute to the Hamiltonian in the local atomic basis, we have
\begin{align}
    \begin{aligned}
        H_{\rm el} &= \sum_n^{N-1} J(\ket{\phi_n}\bra{\phi_{n+1}} + h.c.)\\
        H_{\rm nu} &= \sum_{\alpha=1} \frac{p_{\alpha}^2}{2} + \frac{1}{2}\omega_\alpha^2(q_\alpha^\parallel)^2 + + \frac{1}{2}\omega_\alpha^2(q_\alpha^\perp)^2\\
        H_{\rm el-nu} &= \sum_{n=1}^N \ket{\phi_n}\bra{\phi_n} \sum_\alpha \left( V_{\alpha, \parallel}^n  q_\alpha^\parallel + V_{\alpha, \perp}^n q_\alpha^\perp\right)
    \end{aligned}
\end{align}
where the transverse and longitudinal phonons are summed over separately, and the electron-phonon coupling strength is given by
\begin{equation}
    V_{\alpha, \parallel}^n = \sum_{\mu=1}^N E_{\mu x}^\alpha V_{\mu x}^n, \hspace{0.5 cm} V_{\alpha, \perp}^n = \sum_{\mu=1}^N E_{\mu z}^\alpha V_{\mu z}^n.
\end{equation}

\section{Effects of Removing the Inter-Dot Modes}

\begin{figure}[ht!]
    \centering
    \includegraphics[width=14cm]{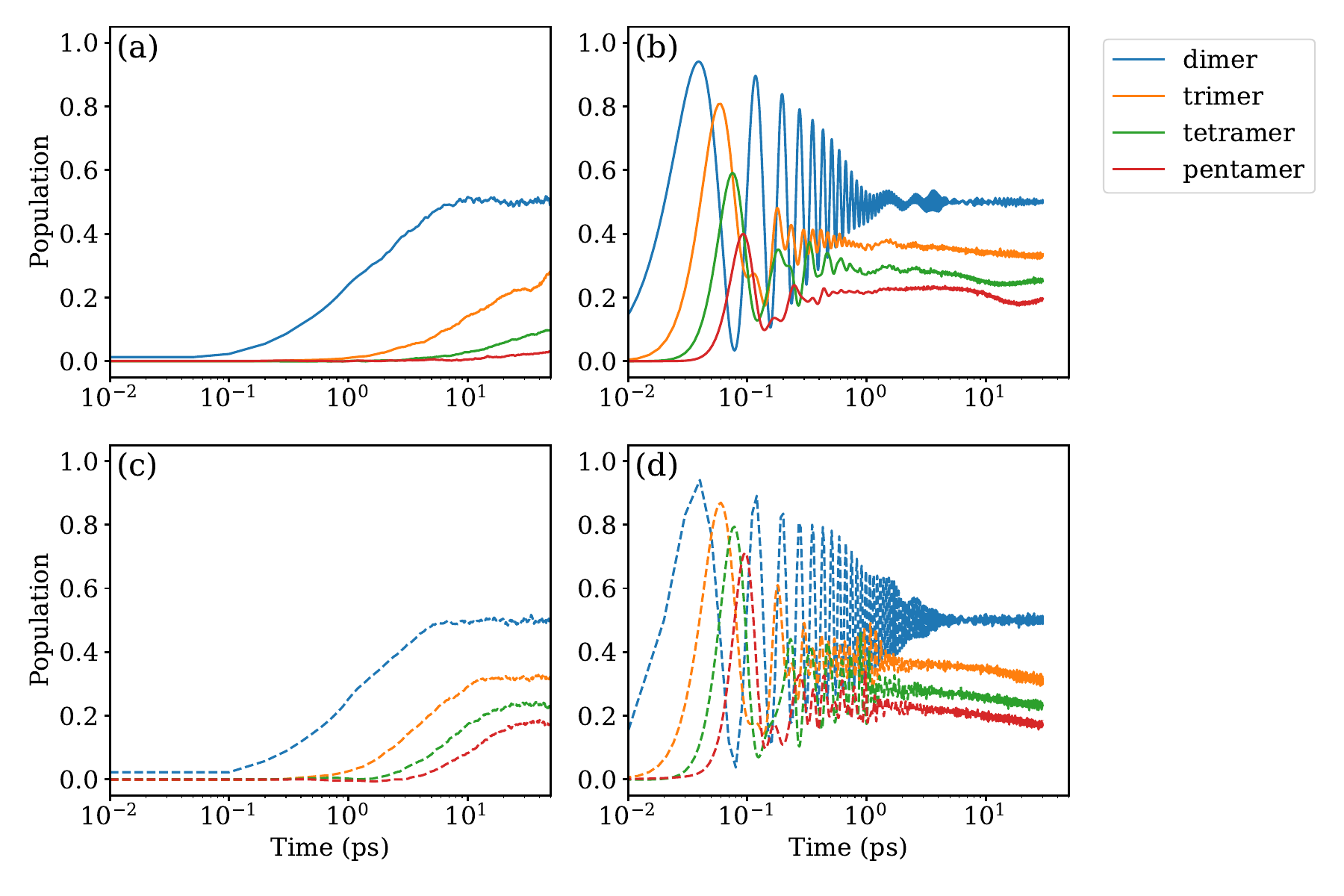}
    \caption{WZ [100] structures in the nonadiabatic and adiabatic electron transfer limits: (a-b) Full modes dynamics (c-d) dynamics with inter-dot modes removed. }
    \label{fig:100_remove_modes_compare}
\end{figure}

\begin{figure}[ht!]
    \centering
    \includegraphics[width=14cm]{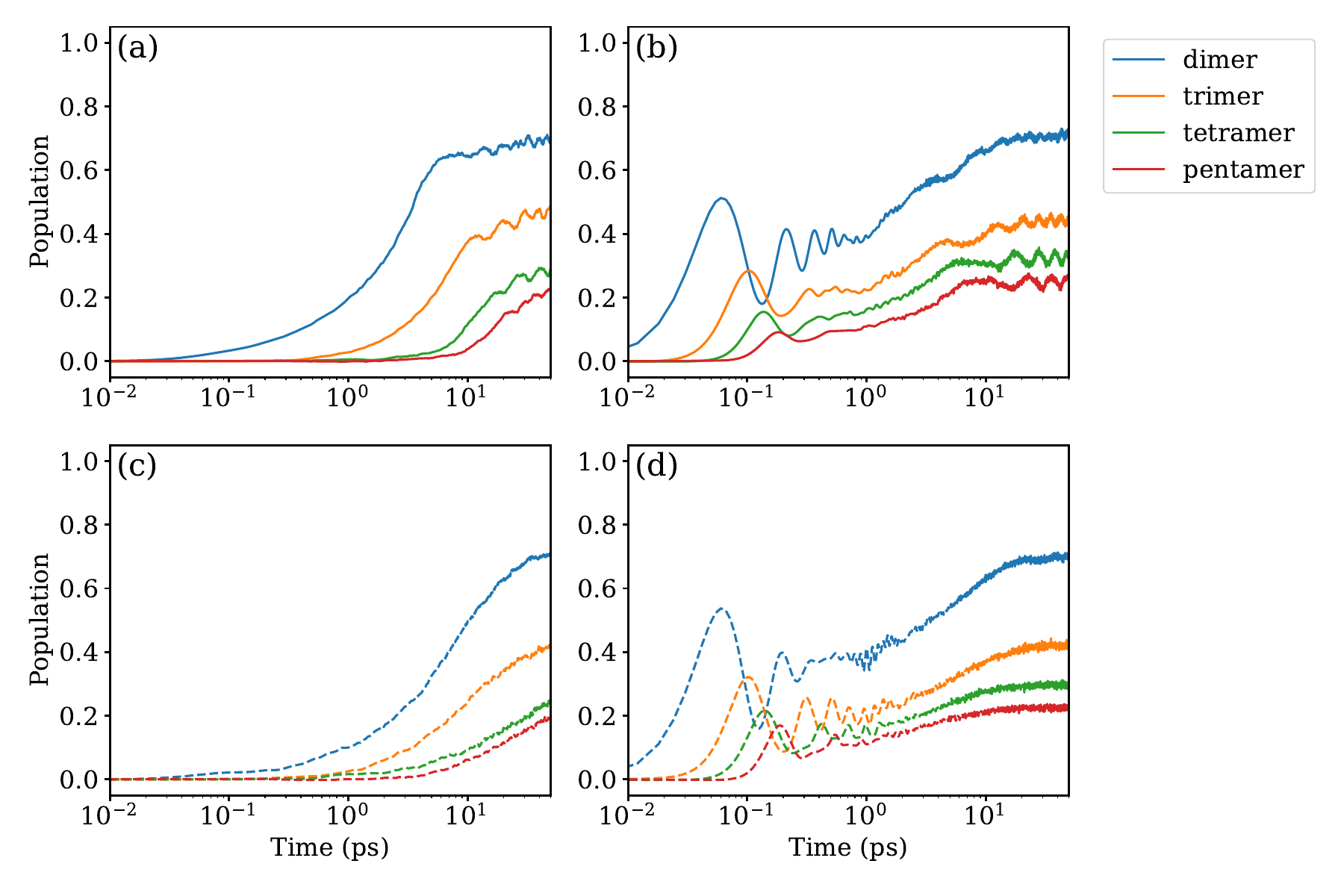}
    \caption{[001] WZ structures in the nonadiabatic and adiabatic electron transfer limits: (a-b) Full modes dynamics (c-d) dynamics with inter-dot modes removed.}
    \label{fig:001_remove_modes_compare}
\end{figure}

\begin{figure}[ht!]
    \centering
    \includegraphics[width=10cm]{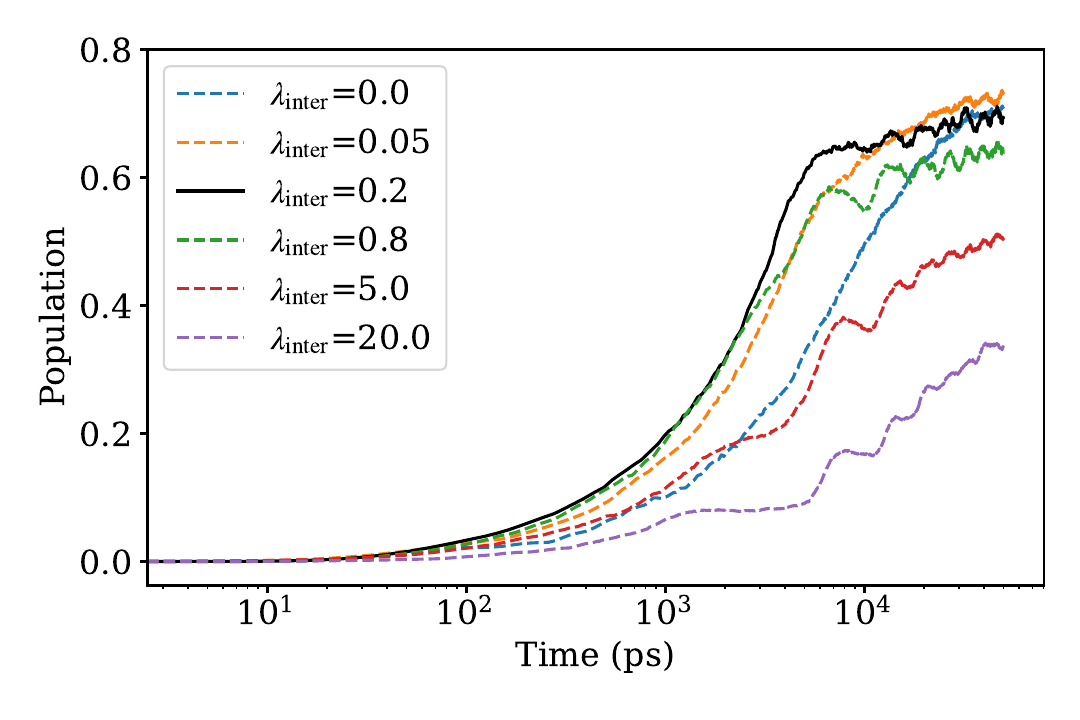}
    \caption{Acceptor population for the WZ [001] dimer as a function of inter-dot reorganization energies $\lambda_{\rm inter}$ (in the unit of meV). The black solid line shows the original reorganization energy $\lambda_{\rm inter}=$ 0.2 meV. }
    \label{fig:remove_mode_scale}
\end{figure}

To understand the effects of inter-dot phonon modes on electron dynamics, we explicitly removed the 6$N_{\rm dot}-6$ inter-dot modes and compared the dynamics in different structures and regimes. This should recover the diffusion limit of electron transfer, where the dynamics are governed by neighboring electronic coupling $J$ and intra-dot electron-phonon couplings. Supplementary Figure~\ref{fig:100_remove_modes_compare} panels (c) and (d) plot the acceptor dynamics for the WZ [100] arrays without inter-dot modes in the nonadiabatic and adiabatic regimes respectively. The main contribution of the inter-dot modes in this configuration is the bending motions. Comparing the dynamics in panels (a) and (c), the transfer time to reach equilibrium is shortened by an order of magnitude when the inter-dot modes are removed in the nonadiabatic regime for tetramers and pentamers. On the other hand, removing the inter-dot modes in the adiabatic regime does not slow down the electron transfer as shown in Supplementary Figure~\ref{fig:100_remove_modes_compare} panels (d). The only noticeable change is the increase in dephasing time when the modes are removed.

Supplementary Figure~\ref{fig:001_remove_modes_compare} plots the dynamics for the WZ [001] arrays where the inter-dot motions are governed by the symmetric stretch. Unlike the [100] structures, the transfer time slightly increases in the nonadiabatic regime when the inter-dot modes are removed as shown in panel (c). The increase in transfer time in the WZ [001] QD arrays is further illustrated in Supplementary Figure~\ref{fig:remove_mode_scale}, where the couplings of inter-dot vibrations are scaled.The transfer time is non-monotonic, first decreasing and then increasing with the rise in inter-dot reorganization energies $\lambda_{\rm inter}$. In the adiabatic regime, removing the modes only affects long-time dynamics, where the oscillation due to inter-dot phonons disappears after 10 ps.

\bibliographystyle{naturemag}
\bibliography{supplement}